\documentclass[prl,twocolumn,showpacs,longbibliography, floatfix]{revtex4-1}
\usepackage{graphicx}
\usepackage{color}
\usepackage{amssymb,amsmath}
\usepackage{bm}
\usepackage[applemac]{inputenc}

\newcommand{\sect}[1]{\textit{#1.---}\ignorespaces}
\newcommand{\degree}{\ensuremath{^\circ}}

\begin{document}

\title{Stacking boundaries and transport in bilayer graphene}

\author{P. San-Jose$^{1*}$, R. V. Gorbachev$^2$, A. K. Geim$^2$, K. S. Novoselov$^4$, F. Guinea$^{1,3}$}
\affiliation{$^1$Instituto de Ciencia de Materiales de Madrid (ICMM-CSIC), Cantoblanco, 28049 Madrid, Spain\\
$^2$Manchester Centre for Mesoscience and Nanotechnology, University of Manchester, Oxford Road, Manchester, M13 9PL, UK\\
$^3$Donostia International Physics Center. (DIPC), P. de Manuel Lardizabal 4, 20018, San Sebasti\'an, Basque Country, Spain. \\
$^4$School of Physics and Astronomy, University of Manchester, Oxford Road, Manchester, M13 9PL, UK}

\begin{abstract}
Pristine bilayer graphene behaves in some instances as an insulator with a transport gap of a few meV.
This behaviour has been interpreted as the result of an intrinsic electronic instability induced by many-body correlations.
Intriguingly, however, some samples of similar mobility exhibit good metallic properties, with a minimal conductivity of the order of $2e^2/h$. Here we propose an explanation for this dichotomy, which is unrelated to electron interactions and based instead on the reversible formation of boundaries between stacking domains (`solitons'). We argue, using a numerical analysis, that the hallmark features of the previously inferred many-body insulating state can be explained by scattering on boundaries between domains with different stacking order (AB and BA). We furthermore present experimental evidence, reinforcing our interpretation, of reversible switching between a metallic and an insulating regime in suspended bilayers when subjected to thermal cycling or high current annealing.
\end{abstract}

\date{\today}
\maketitle

Bilayer graphene presents an interesting case in terms of its electronic properties at low energy \cite{Neto:RMP09}. Its bandstructure around the K and K' points displays hyperbolic bands touching at the neutrality point, if trigonal warping terms  \cite{McCann:PRL06a,Kechedzhi:PRL07} are neglected. The low-energy electronic structure, high density of states (DOS) and high level of degeneracy in a magnetic field \cite{Novoselov:NP06,McCann:PRL06a} can potentially lead to multiple competing broken symmetry states both in zero and finite magnetic fields \cite{Ezawa:JPSJ07,Castro:PRL08,Barlas:PRL08,Abanin:PRL09,Vafek:PRB10a,Zhao:PRL10a,Lemonik:PRB10,Zhang:PRB10a,Martin:PRL10,Vafek:PRB10,Jung:PRB11,Zhang:PRL11,Zhang:PRL12,Kharitonov:PRB12}. Such many-body instabilities may polarise pseudospin and layer quantum numbers, break the hexagonal symmetry, and possibly also open spectral gaps. Furthermore, it has been shown that at low enough energies, the electronic structure becomes more complex: trigonal warping leads to appearance of four linear cones around each of the K (K') points \cite{McCann:PRL06a}. The suppressed density of states in such case would lead to a very different types of instabilities than for a hyperbolic spectrum \cite{Mayorov:S11,Vafek:PRB10}.

The experimental evidence for some of the predicted instabilities is still debated. Particularly controversial is the basic question of whether pristine graphene bilayers exhibits a spectral gap around neutrality in the absence of a magnetic field. Some experimental groups have reported a metallic, and hence gapless, ground state, while others find their cleanest samples to be insulating. Among the former are reports of  bilayers on hexagonal boron nitride \cite{Maher:NP13} and suspended graphene \cite{Feldman:NP09,Weitz:S10} samples. Other groups reported a suppression of the DOS in zero magnetic fields, arguably due to symmetry breaking, although again no insulating behavior was observed \cite{Martin:PRL10,Mayorov:S11}.

Conversely, two groups reported the observations of insulating behaviour in some of their samples \cite{Bao:PNAS12,Velasco:NN12,Freitag:PRL12,Freitag:SSC12,Freitag:PRB13}, with transport gaps of about $\sim 2$ meV, even in zero magnetic field. These samples are suspended \cite{Du:NN08,Bolotin:SSC08}, and have high mobility as a result of  current annealing with  high currents  \cite{Bolotin:SSC08}, a process that heats up the samples above 1000\degree C \cite{Dorgan:NL13}. Notably, however, both groups also find some of their high-mobility  samples to be metallic \cite{Bao:PNAS12,Velasco:NN12,Freitag:PRL12}. A tentative explanation for this dichotomy was proposed \cite{Bao:PNAS12}, whereby the true bulk ground state is insulating, but the coexistence of two different broken symmetry domains leads to transport along domain walls bridging the contacts.

\begin{figure}
   \centering
   \includegraphics[width=0.7\columnwidth]{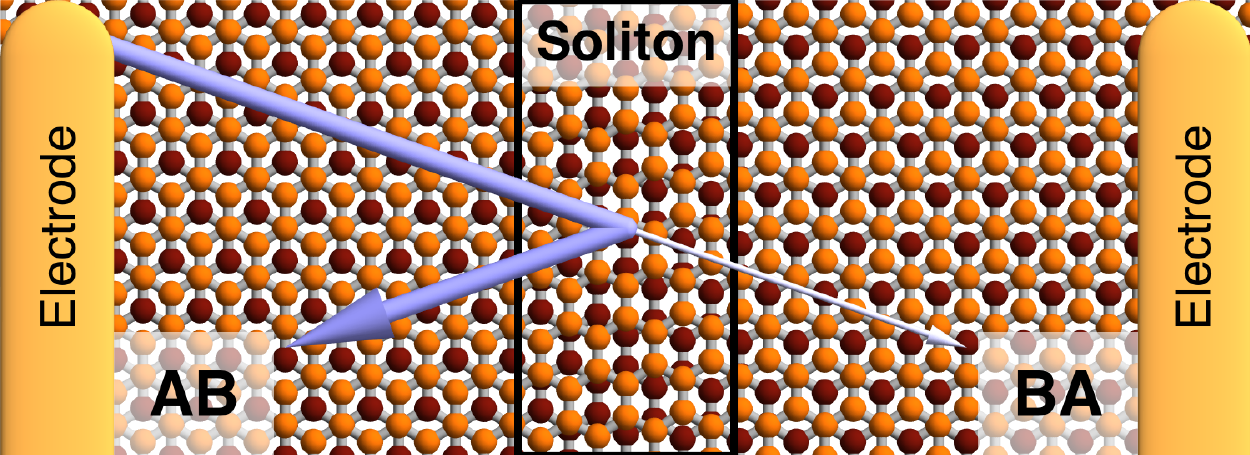}
   \caption{A smooth stacking boundary (soliton) between AB- and BA-stacked graphene bilayers may act as a low transparency barrier for low energy electrons.}
    \label{fig:tocimage}
\end{figure}

\begin{figure}
   \centering
   \includegraphics[width=\columnwidth]{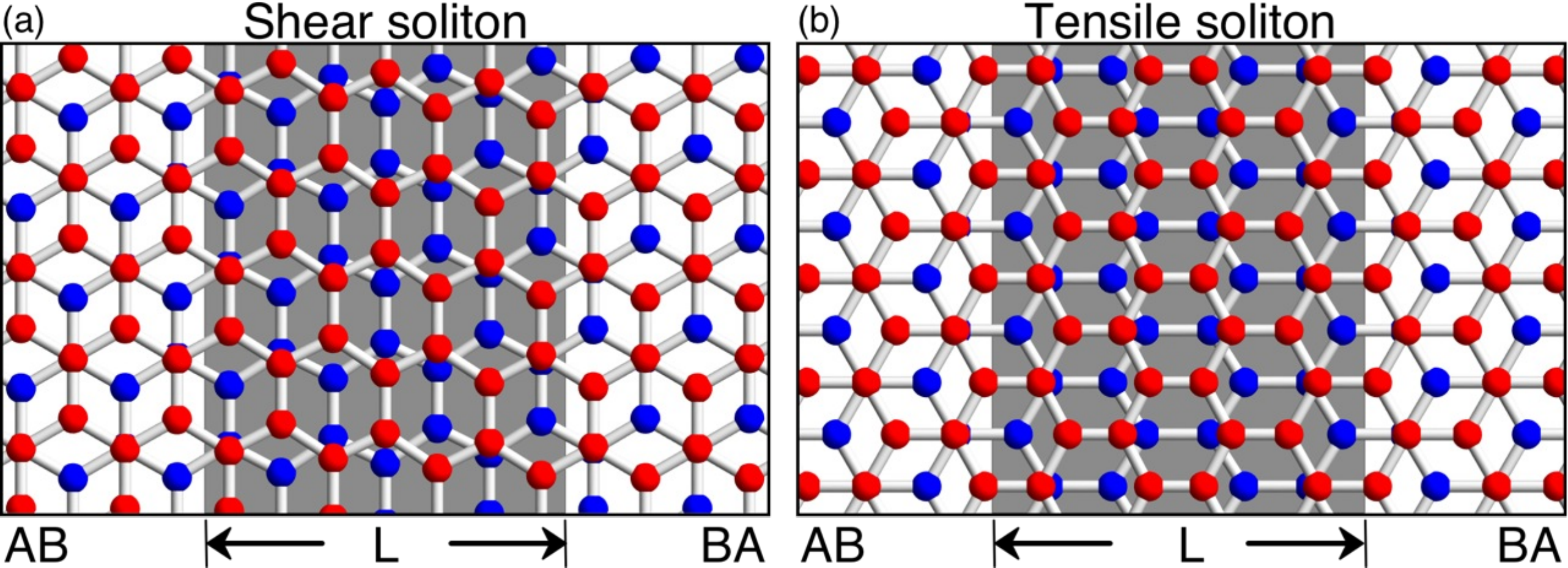}
   \caption{Crystalline structure of a stacking boundary of the shear type (a) and the tensile type (b). $L$ is the thickness, red and blue are atoms in the top and bottom layers, respectively.}
    \label{fig:solitons}
\end{figure}

In this work we propose an alternative interpretation, which does not require a spectral gap in the bulk, and is based instead on the presence of stacking domain walls, recently observed in bilayers \cite{Lin:NL13,Alden:PNAS13}. Stacking domain walls, also known as stacking solitons, are boundaries between regions of well defined AB and BA stackings, see Fig. \ref{fig:solitons}. They have been found to be ubiquitous in bilayers, particularly when heated to temperatures above 1000\degree C, at which point solitons become mobile \cite{Alden:PNAS13}, and may emerge spontaneously. As most of the high-quality bilayer samples are free-standing and current-annealed \cite{Bao:PNAS12,Velasco:NN12,Freitag:PRL12,Freitag:SSC12,Freitag:PRB13} to temperatures above this threshold \cite{Dorgan:NL13}, the formation of stacking defects is extremely plausible, and must therefore be taken into account when interpreting experimental data for current-annealed bilayers. We find that their presence may strongly suppress transport, and result in an insulating behaviour of an otherwise metallic bilayer.

Calculations of transport through stacking boundaries in graphene bilayers was  studied in Ref. \cite{Koshino:PRB13}. Our numerical simulations extend these results, showing that scattering on stacking boundaries across a device may give rise to transport gaps in the $1-5~\mathrm{meV}$ range, and exhibit features in the differential conductance profile compatible with those observed in insulating bilayers \cite{Bao:PNAS12,Velasco:NN12,Freitag:PRL12,Freitag:SSC12,Freitag:PRB13}. Supporting this interpretation, we furthermore present experimental transport results that reveals the reversible nature of the metallic and insulating regimes. We find that suspended bilayer samples can be switched repeatedly between metallic and insulating by applying high annealing currents and thermal cycling, as expected from the formation of stacking boundaries across the sample.
We speculate that this type of defects, and not many-body instabilities, may underlie the observations of Refs. \cite{Bao:PNAS12,Velasco:NN12,Freitag:PRL12,Freitag:SSC12,Freitag:PRB13}


\sect{Numerical results}
A stacking soliton is a smooth boundary between an AB- and a BA-stacked bilayer region. In the simplest, lowest energy configuration, such boundary is straight, and connects two points on the edges of the bilayer. The precise profile, orientation and thickness of a soliton depend on the stress field that creates it. A relative interlayer shear produces a shear soliton, while a relative uniaxial strain results in a tensile soliton,  see Fig. \ref{fig:solitons}. All intermediate configurations between shear and tensile solitons are possible, depending on the orientation of the soliton with respect to the underlying lattice, although the shear boundary has the least energy. All such configurations are minimal AB/BA boundaries, with saddle-point (SP) stacking at the center of the soliton \cite{Alden:PNAS13}. Other possibilities, such as AB/AA/BA configurations are possible \cite{Koshino:PRB13}, but are seldom observed in real samples \cite{Alden:PNAS13} due to their higher energy density.

Electron transport across a soliton may be computed using a tight-binding approach or, more efficiently, using a low energy Dirac fermion model in the continuum, following Koshino \cite{Koshino:PRB13}. The soliton stacking profile is incorporated into a position-dependent interlayer coupling (see Supplementary Information for details on the model). For the case of zero magnetic field, we take the latter approach, neglecting trigonal warping terms for simplicity. Valley mixing can also be neglected for a realistic soliton thickness. We employ a recursive Green's function method \cite{Datta:97} to calculate the transmission $T(\epsilon, k_y)$ of an electron incident at energy $\epsilon$ with a (conserved) momentum $\hbar k_y$ parallel to the soliton, which is assumed straight and aligned along the $y$ direction (see Supplementary Information). The differential conductance across a non-interacting sample is expressed as $dI/dV=g_s g_v \frac{e^2}{h} T(E_F+eV)$, where the integrated transmission is $T(\epsilon)=W\int \frac{dk_y}{2\pi} T(\epsilon, k_y)$, and $g_s=g_v=2$ are the spin and valley degeneracies respectively.
$W$ is the sample width, $L$ is the soliton thickness (defined as the FWHM of the minimum interlayer hopping), and $E_F$ is the Fermi energy. We will focus on the  neutrality point $E_F=0$ with vanishing equilibrium carrier density $n$.

\begin{figure}
   \centering
   \includegraphics[width=0.49\columnwidth]{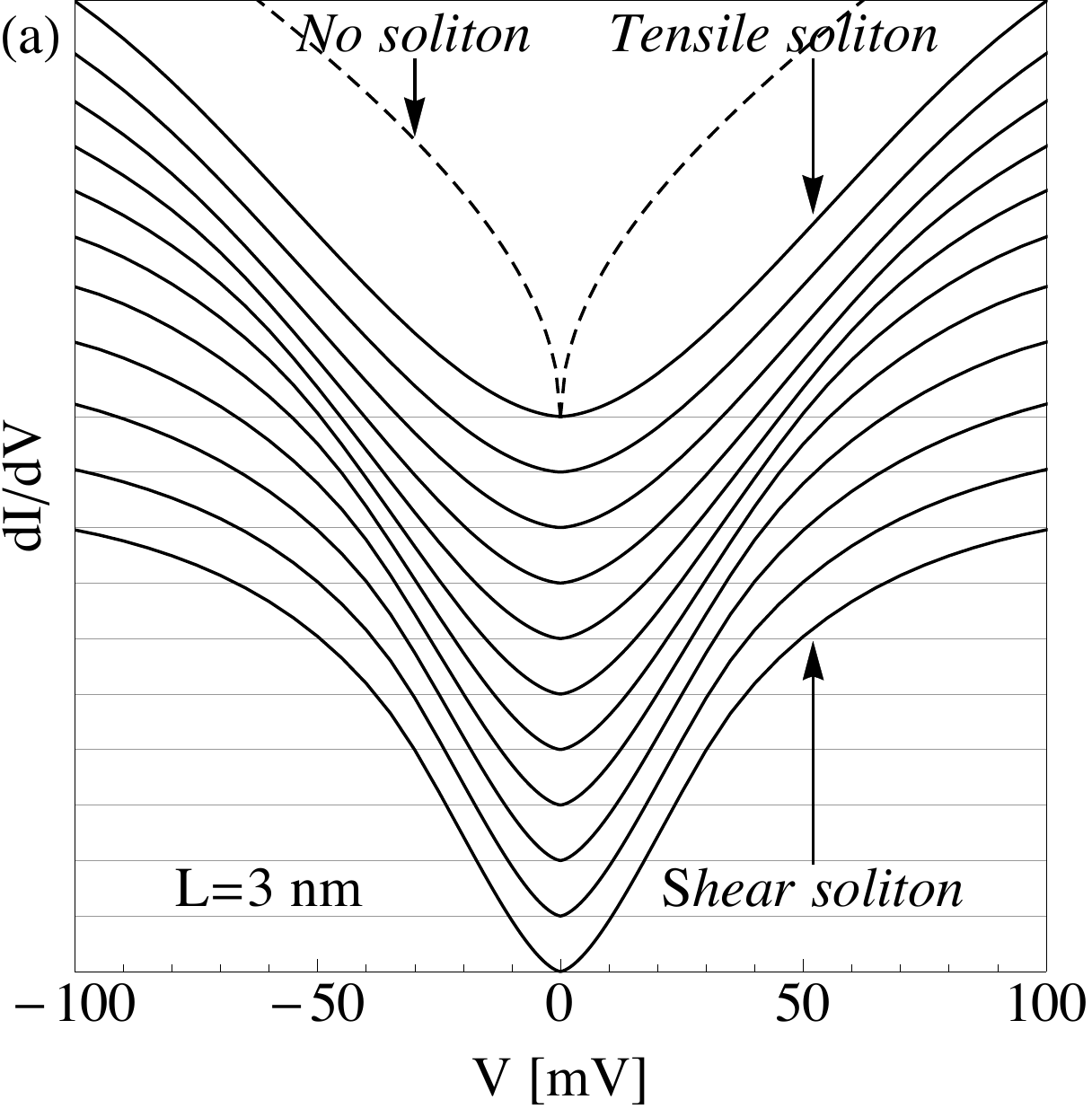}
   \includegraphics[width=0.49\columnwidth]{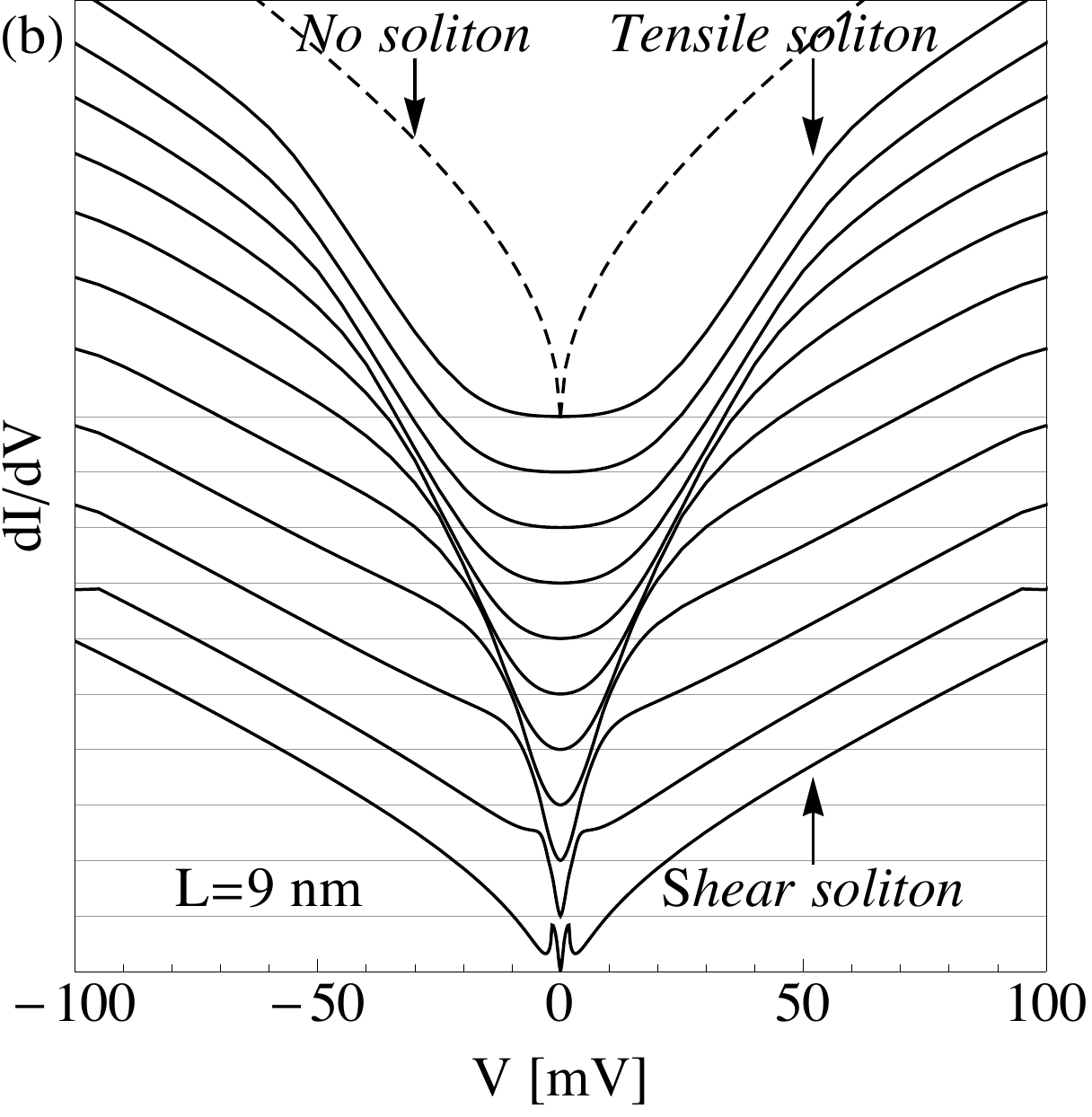}
   \caption{Differential conductance  versus bias across a soliton in neutral graphene for various soliton orientations, ranging from a shear to a tensile soliton (curves are offset for clarity).  Panel (a) corresponds to a soliton thickness $L=3$ nm, and (b) to $L=9$ nm. The thicker shear solitons exhibit low bias features at both sides of the transport gap. The dashed line on the tensile offset corresponds to the metallic $dI/dV$ without a soliton.}
   \label{fig:Ttheta}
\end{figure}

\begin{figure}
   \centering
   \includegraphics[height=0.5\columnwidth]{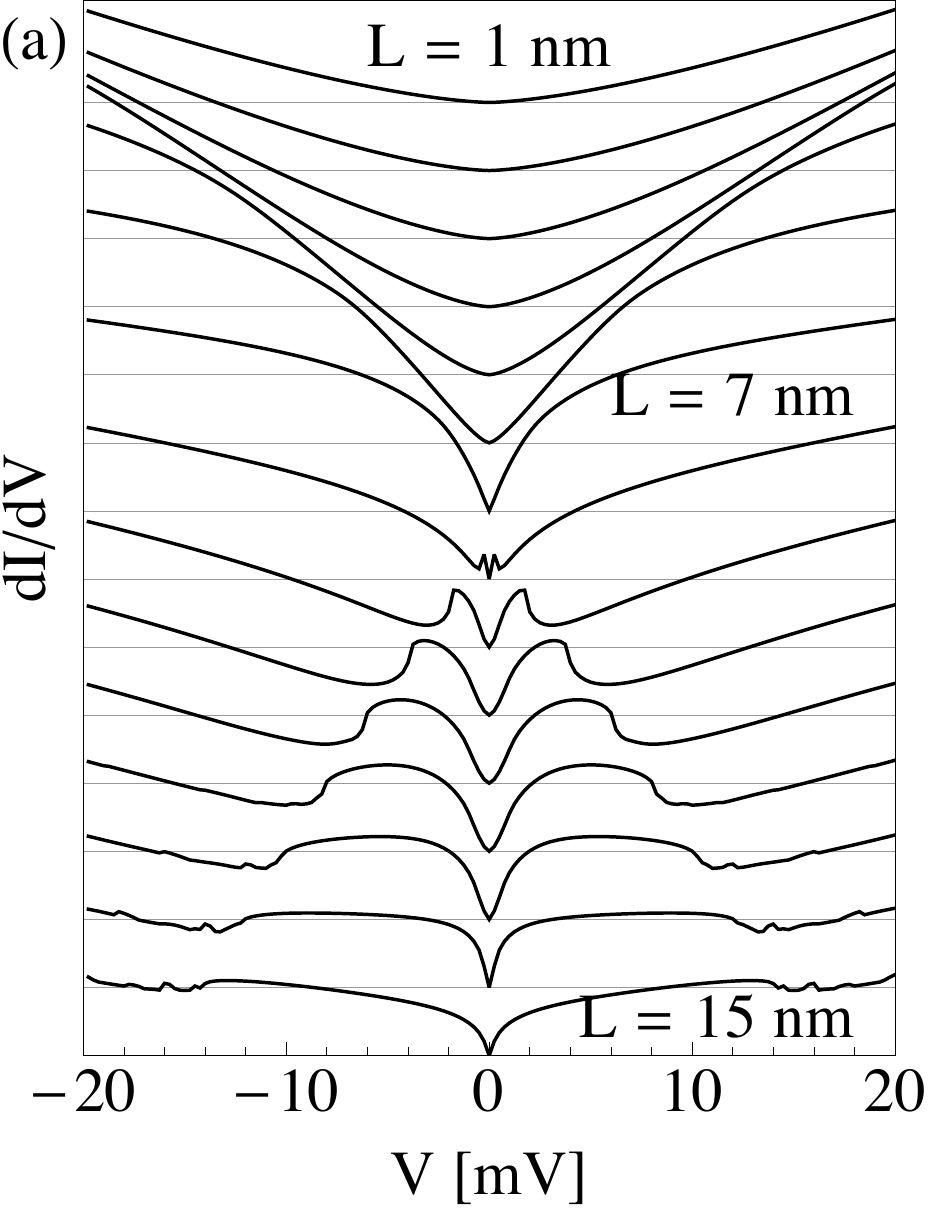}
   \includegraphics[height=0.5\columnwidth]{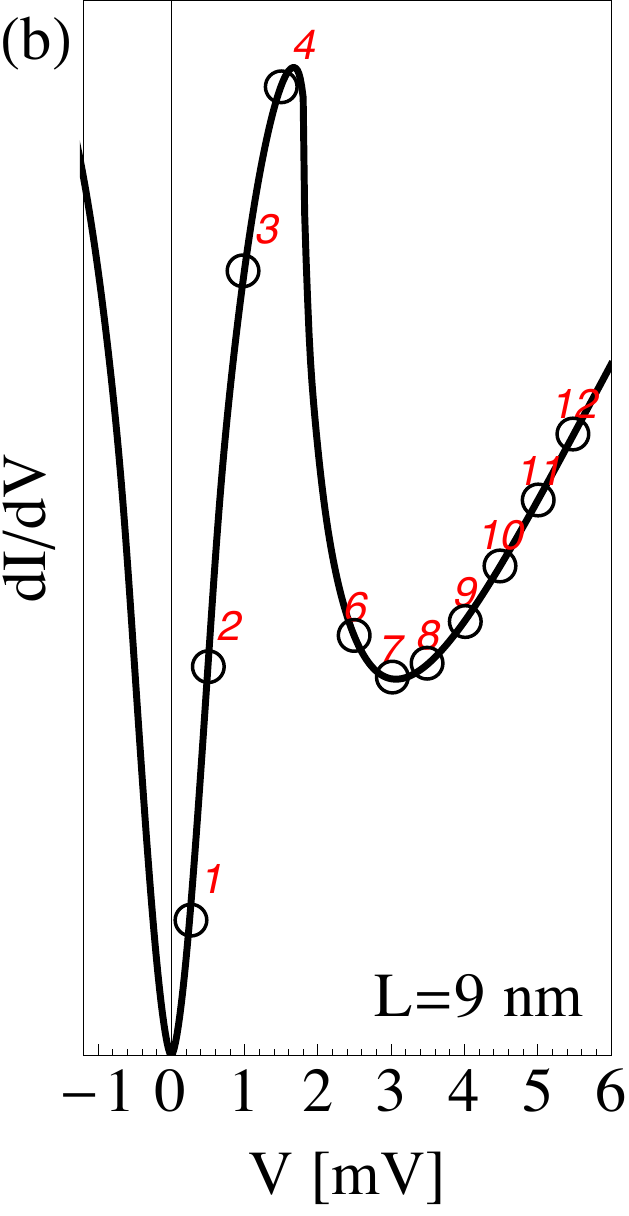}
   \includegraphics[height=0.5\columnwidth]{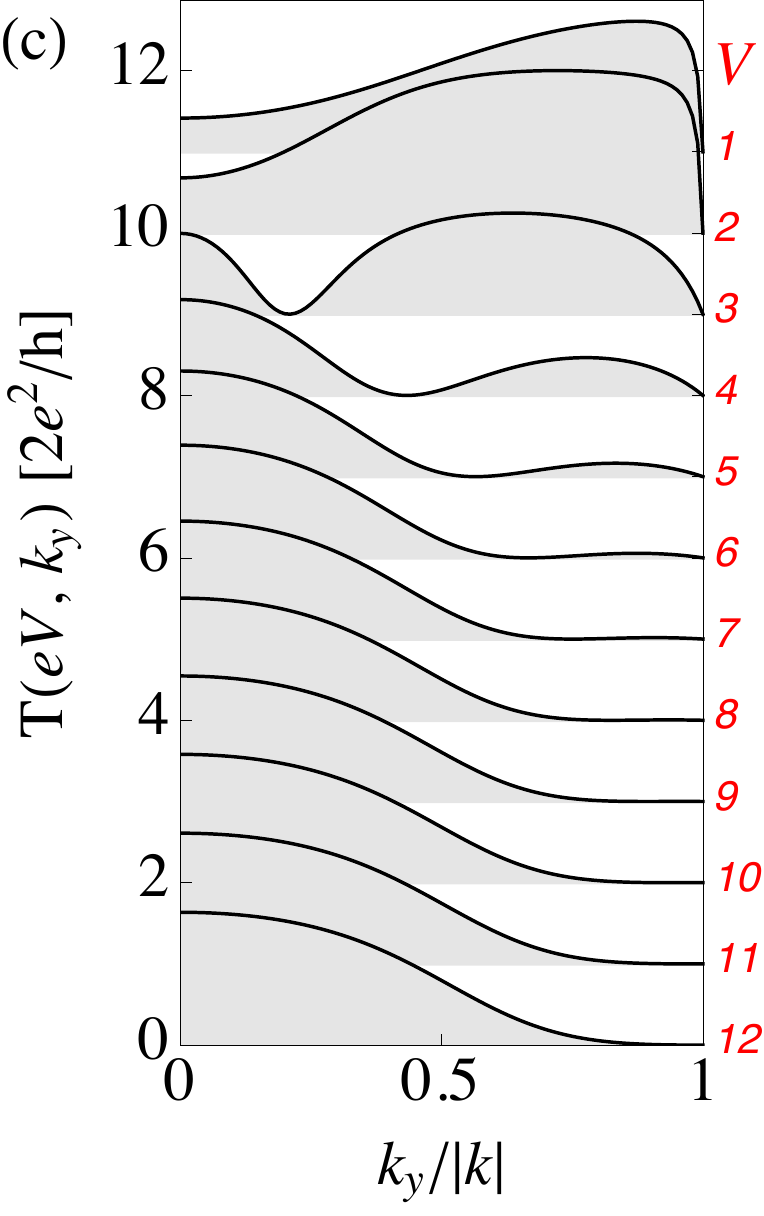}
   \caption{(a) Differential conductance in neutral graphene across a shear soliton for increasing soliton thickness $L$ (curves from top to bottom, offset for clarity). A non-monotonous feature arises around zero bias voltage for thickness $L\gtrsim 7$ nm. (b) Blow-up of the $L=9$ nm conductance. (c) Transmission resolved in transverse momentum (offset for clarity), corresponding to numbered circles in (b), and exhibiting a transition from grazing angle- (low bias) into normal incidence-dominated transmission (large bias) across the peaks in (b).}
   \label{fig:T}
\end{figure}

In Fig. \ref{fig:Ttheta} we present the differential conductance versus bias in neutral graphene, for two values of soliton thickness $L$. Different curves correspond to different soliton orientations ranging from shear to tensile. The opposite chirality of AB and BA regions \cite{Novoselov:NP06} results in a strong suppression of the conductance around zero bias, particularly in the case of the tensile soliton.
The bias window exhibiting this insulating behaviour grows wider and better defined, for a tensile soliton, as its thickness $L$ is increased, compare top curves in Figs. \ref{fig:Ttheta}(a,b). In contrast, the shear soliton displays a decreasing soft transport gap as $L$ increases, and a non-monotonous $dI/dV$ profile above a thickness $L\sim 7~\mathrm{nm}\approx l^\mathrm{SP}_\perp$ (where $l^\mathrm{SP}_\perp=\hbar v_F/\gamma_\mathrm{SP}\approx 6.6$ nm is the interlayer coupling length at the centre of the soliton \cite{Snyman:PRB07}), see Fig. \ref{fig:T}(a,b). This profile is characterised by a narrow U-shaped transport gap at zero bias, with  two side-peaks at around $V\approx \pm(1 - 5~\mathrm{mV})$, depending on $L$. Such structure is strongly reminiscent of the differential conductance measured in Refs. \cite{Bao:PNAS12,Velasco:NN12,Freitag:PRL12,Freitag:SSC12,Freitag:PRB13}, although its origin is not a many-body bulk instability, but rather a transition between two different single-particle transport regimes. At low bias (between the peaks), $k L<1$ (where $\hbar k$ is the total momentum of incoming carriers), so the soliton behaves as an abrupt barrier between regions of opposite carrier chirality. Hence, transmission is minimum for normal incidence $k_y=0$ \cite{Katsnelson:NP06}, and is maximized at grazing angles $k_y\sim k$, see Fig. \ref{fig:T}(c). For higher energies, the soliton appears as an adiabatic barrier, since $L$ is greater than the wavelength, and also greater than the interlayer coupling length $l_\perp^\mathrm{SP}$. The transmission pattern is then the opposite, with a maximum at normal incidence. The peak in the differential conductance corresponds to the transition between the two regimes, wherein transmission is high for all incident angles. We found that these transport features persist, and are even enhanced, when a number of solitons are present in the sample, and are not sensitive to distortions or misalignments (see Supplementary Information for more details).

\begin{figure}
   \centering
   \includegraphics[height=0.43\columnwidth]{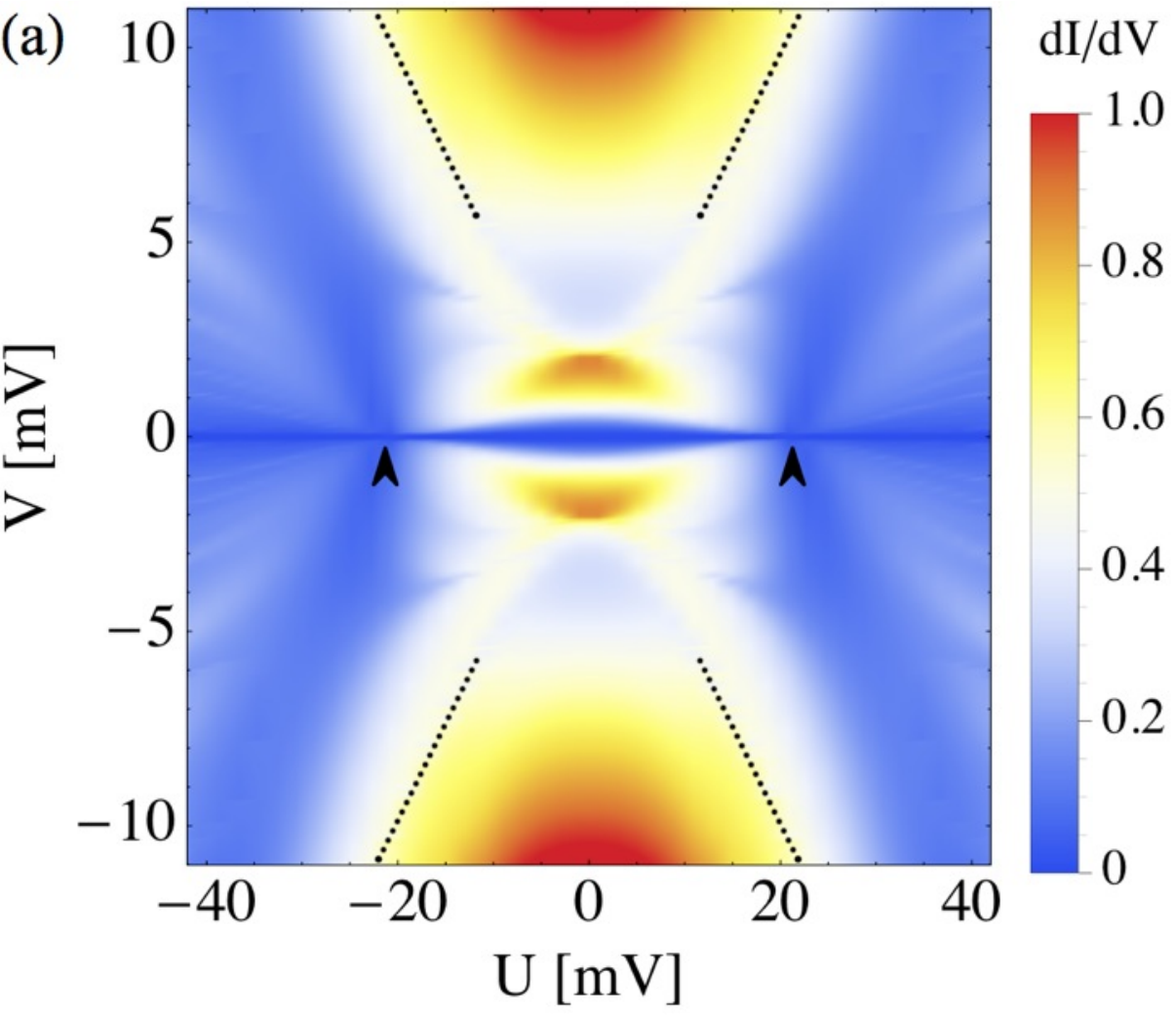}
   \includegraphics[height=0.43\columnwidth]{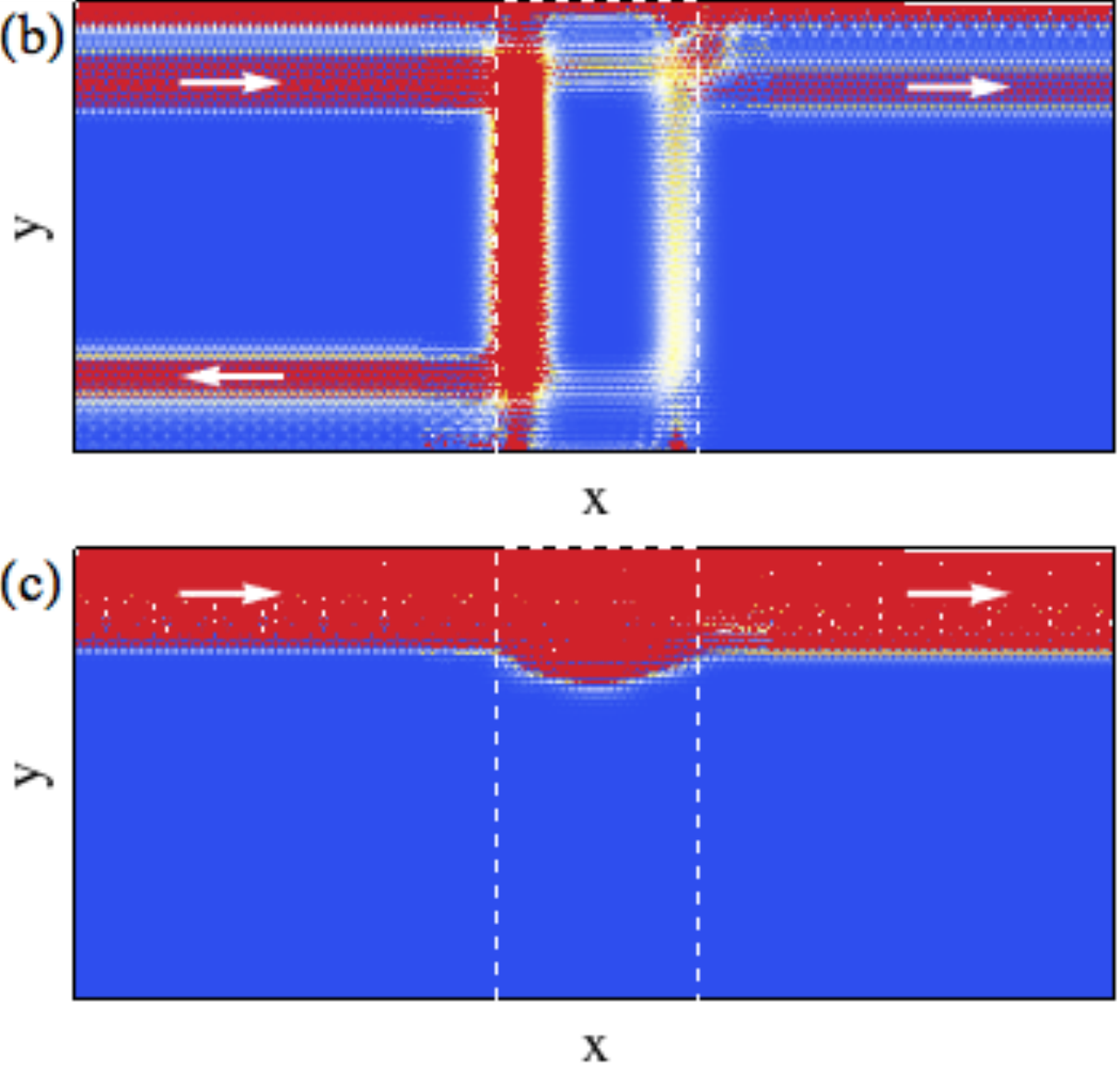}
   \caption{(a) Differential conductance as a function of electrode bias $V$ and interlayer bias $U$, applied across a finite region around a shear soliton. (b,c) Edge mode propagation (arrows indicate direction) across a shear soliton (dashed line) on a finite width bilayer nanoribbon with a uniform magnetic flux. For a small energy (b), the soliton can efficiently connect the two edges, suppressing the quantum Hall effect quantization, while at higher energies (c) the edge mode propagates through the soliton with perfect transparency.}
   \label{fig:TvsU}
\end{figure}

An interlayer voltage $U$, as created by a backgate-topgate arrangement on the bilayer,  modifies the differential conductance in a way once more reminiscent of reported observations \cite{Velasco:NN12,Freitag:PRL12,Freitag:SSC12}, see Fig. \ref{fig:TvsU}(a). The transport gap and side peaks around $V=0$ vanish as the interlayer voltage is increased (black arrowheads), due to a breaking of chirality that underlies these features. At high enough $U$, the transport gap reopens, as a result of the spectral gap of magnitude $U$ in the bulk (dotted lines) \cite{Castro:PRL07,Oostinga:NM08}.
(Note that, additionally, topologically protected modes confined to the soliton will arise under a finite bias $U$ \cite{Wright:APL11,Vaezi:PRX13,Zhang:PNAS13, San-Jose:PRB13}). 
All these features are once more strongly reminiscent of existing measurements \cite{Velasco:NN12,Freitag:PRL12}.

A uniform magnetic flux also suppresses transport at low energies. To compute this effect, it becomes necessary to use a finite width ribbon, best modelled in a tight-binding approximation. The total transmission in this case is a sum over the transmission of open modes which, for completely filled Landau levels, are tightly confined to the edges. A stacking soliton across the sample can potentially destroy the Hall effect quantisation by inducing strong interedge scattering, since it is an extended defect connecting opposite edges. Such increase in the magnetoresistance is confirmed by our simulations at low enough energies, such that the soliton appears abrupt on the scale of the edge mode wavelength. Above a threshold energy, the soliton appears adiabatic once again, edge modes remain decoupled, and quantum Hall quantisation is restored (see Supplementary Information for more details).  

While this single-particle picture of the insulating state can account for a wide range of transport observations in suspended bilayers, and is simpler than the alternative explanation based on many-body instabilities, it is not immediately clear whether some evidence exists that allows us to distinguish between the two interpretations.
One proposed explanation for the metallic state with a gapped bulk is that edge states may become confined along a domain wall that short-circuits the contacts.  These states would contribute to a finite conductance in a bilayer that is otherwise gapped by many body effects \cite{Bao:PNAS12}. (A similar situation arises if a boundary between regions with gaps of opposite signs connects the leads \cite{Wright:APL11,Vaezi:PRX13,Zhang:PNAS13,San-Jose:PRB13}).
However, ballistic edge channels would only contribute to conductance with a fixed integer number of quanta $e^2/h$, which would yield a conductivity dependence on sample size. This is at odds with a rather universal minimal conductivity $\sim 2-3 e^2/h$ observed in metallic bilayers \cite{Bao:PNAS12}. This scenario would also imply the emergence of a plateau in the conductivity versus Fermi energy, up to energies of the order of the transport gap, again seemingly incompatible with the observations \cite{Bao:PNAS12,Velasco:NN12,Freitag:PRL12,Freitag:SSC12,Freitag:PRB13}. Lastly, any atomically sharp defect in the bilayer would mix valleys, and thus induce backscattering and localisation in the edge states, since topological protection is confined to each valley. Thus, it is very unlikely that in real samples such solitons would support ballistic transport to provide conductivity $\sim 2-3 e^2/h$.

\begin{figure}
   \centering
   \includegraphics[height=1.2\columnwidth]{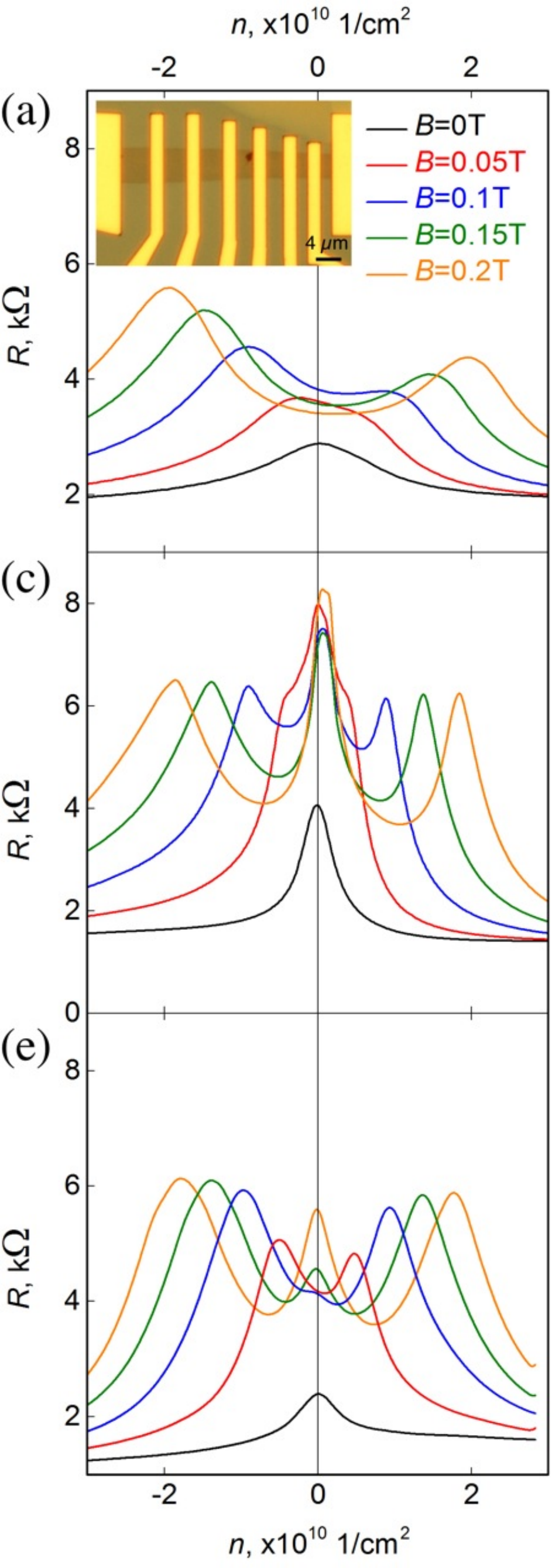}\hspace{0.2cm}
   \includegraphics[height=1.2\columnwidth]{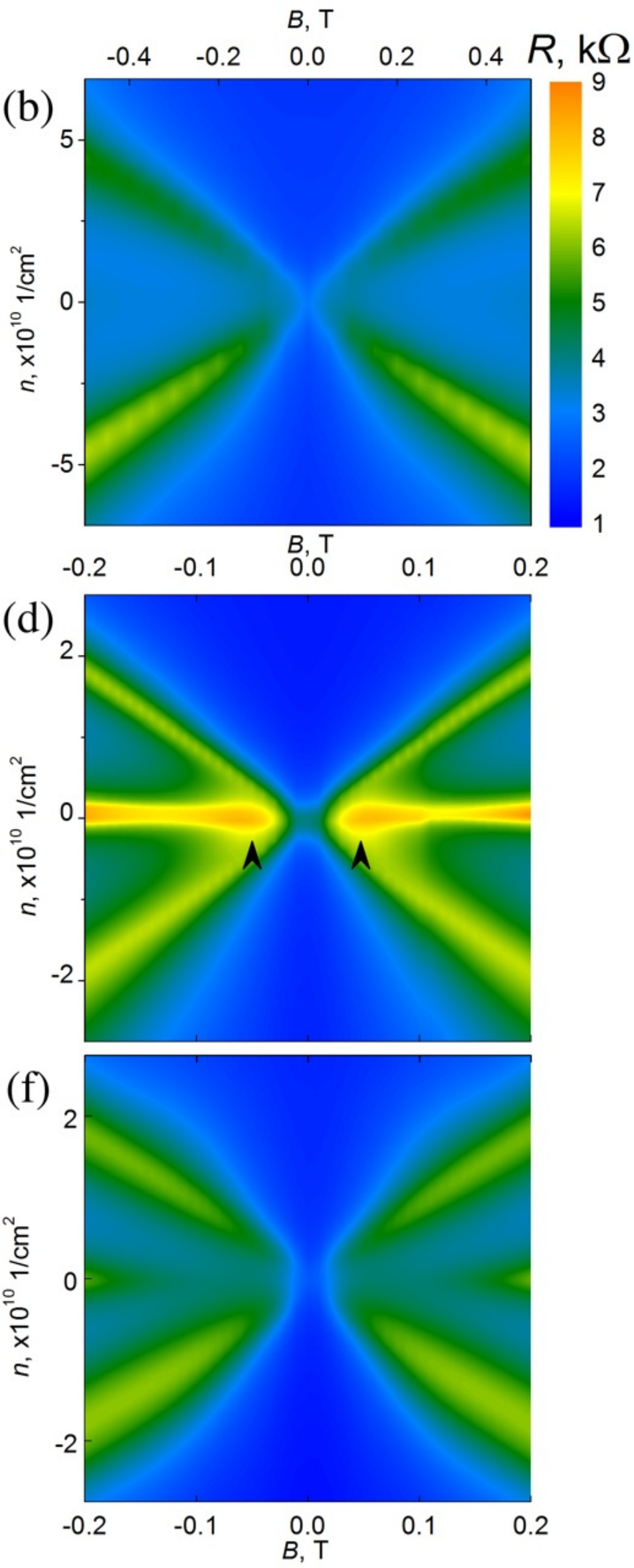}
    \caption{Dependence of resistance of one of our devices on carrier density $n$ and magnetic field $B$. The three rows of panels represent different stages of annealing, with the top row (a,b) -- before the switching to the insulating state; middle row (c,d) -- the insulating stage (note the enhanced low field magnetoresistance - arrowheads); and bottom row (e,f) -- after thermal recycling and additional annealing. Inset in (a), micrograph of one of our samples.}
   \label{fig:R}
\end{figure}

\sect{Experimental results}
We now present further evidence in support of the stacking boundary scenario for insulating bilayers.  We have performed two-terminal magnetotransport measurements across suspended graphene bilayers. We have found a reversible switching between distinct metallic and insulating transport regimes, that may be understood in terms of the formation and annealing of stacking boundaries across the sample.
The type of the samples and sample preparation procedure is the same as described in our previous work  \cite{Mayorov:S11}. In brief, narrow (2-4 $\mu$m) graphene stripes were prepared on top of Si/SiO$_2$(300 nm) substrate. A set of Cr/Au (5 nm/150 nm) contacts, inset of Fig. \ref{fig:R}(a), was prepared via e-beam lithography, e-beam evaporation and lift-off to form two-probe devices. 150 nm of SiO$_2$ has been etched away in buffered hydrofluoric acid to form free-standing devices.

As prepared, the samples were usually p-doped and exhibited mobilities of around 5,000 cm$^2$/Vs - typical for devices on silicon oxide \cite{Novoselov:S04}. Current annealing was employed in order to improve the quality of the devices. We gradually increased the annealing current densities $J$, starting at about 0.2 mA per micrometer width of the device. The typical maximum current densities employed, above which the samples usually would burn down, were around $J_\mathrm{max}\approx 1~\mathrm{m A/\mu m}$.

Determining the transport mobility for our two-terminal devices is not trivial, since contact resistance can be considerable. Therefore, we employ quantum mobility $\mu_q$ as a measure of the quality of our samples \cite{Mayorov:S11}. We determine $\mu_q$ from $\mu_q B_0=1$, where $B_0$ is the onset magnetic field for quantum oscillations. A significant improvement in the quality of our samples can be observed already at annealing currents roughly 75\% of $J_\mathrm{max}$.

Figs. \ref{fig:R}(a,b) show the resistance $R$ of a device in such state, with $\mu_q$ of the order of $10^5~\mathrm{cm^2/Vs}$, as a function of  $n$ and $B$.
In zero magnetic field, the device is metallic, with a maximum resistance $R\approx 0.08 R_K$ ($R_K=h/e^2$ is the von Klitzing constant). At this stage of annealing our devices would be typically undoped (within $10^8~ \mathrm{cm}^{-2}$), with the resistance peak situated practically at zero gate voltage. In low magnetic field, the peak splits into two, as expected for a Fermi level situated in the gap of filling factor $\nu=\pm 4$.
The pronounced minima at $n=0$ indicates the absence of degeneracy lifting for the $N=0$ and $N=1$ Landau levels positioned at zero energy \cite{Novoselov:NP06,McCann:PRL06a}. At higher magnetic fields, above a certain threshold $B_\mathrm{th}$ a pronounced peak appears at $n=0$ [not shown on Fig. \ref{fig:R}(a,b)], indicating a lifting of the 8-fold degeneracy of the zero energy Landau level, in accordance with previous literature \cite{Feldman:NP09,Weitz:S10,Maher:NP13,Bao:PNAS12,Martin:PRL10,Mayorov:S11,Velasco:NN12,Freitag:PRL12,Freitag:SSC12, Freitag:PRB13}.

The use of a $J$ higher than 75\% of $J_\mathrm{max}$ often leads to abrupt changes in the electronic properties of the samples. In such a new state, the resistance would increase several times [Fig. \ref{fig:R}(c,d)], though, notably, $\mu_q$ is still of the order of $10^5~\mathrm{cm^2/V s}$.
The most pronounced changes can be observed in low magnetic fields. Here we do not find a minimum at $n=0$ anymore, but rather an insulating state starting from magnetic fields below the many-body threshold $B_\mathrm{th}$ [see features marked by arrowheads in Fig. \ref{fig:R}(d)]. This positive magnetoresistance around $n=0$ is consistent with inter-edge scattering on a soliton [Fig. \ref{fig:TvsU}(b)].

The switching between transport regimes is reversible. By warming up the sample in its insulating state to 300K and cooling down again, the original, non-insulating state [as presented on Fig. \ref{fig:R}(a,b)] can be restored. Such re-enterant behaviour can be observed several times on the same sample. Occasionally, by careful selection of $J$, it is possible to anneal our devices to even cleaner states \cite{Mayorov:S11,Mayorov:NL12} with mobilities up to $10^6~\mathrm{cm^2/V s}$, Fig. \ref{fig:R}(e,f). Again, in such state, a dip in the two probe resistance is observed at low magnetic fields, which eventually develops into a broken symmetry state above a threshold $B_\mathrm{th}$. 

Although the above re-entrant behaviour has not been observed previously, the pattern of obtaining two types of samples with similar mobilities  -- insulating and conducting ones -- is  familiar from the previous work of other groups \cite{Bao:PNAS12,Velasco:NN12,Freitag:PRL12,Freitag:SSC12,Freitag:PRB13}. Furthermore, typical mobilities achieved in those experiments are similar to what is described in this report. Thus, we speculate that all these observations   have the same origin, and point to gapless clean bilayers that develop stacking boundaries during high temperature annealing. The possibility of moving, creating and annihilating these boundaries at such temperatures \cite{Alden:PNAS13} may explain the observation of both insulating and metallic regimes in similar samples.

 The work reported here shows that scattering on stacking AB/BA boundaries results in a generic insulating-like behaviour of an otherwise metallic graphene bilayer. This is a consequence of the special chiral properties of charge carriers in bilayer graphene. A reversible thermal switching between insulating and metallic regimes can also be induced in clean samples. Our experimental observations, and also a number of measurements from other groups in graphene bilayers near the neutrality point, are consistent with the presented model for transport anomalies across stacking solitons. Further work is required to provide direct evidence for correlations between stacking boundary arrangements and apparent energy gaps. However, it is clear from our analysis that such boundaries strongly influence transport properties of bilayer graphene and, therefore, should be ruled out before alternative, many-body models are invoked. The modification of the transport properties of graphene by creating and manipulating extended defects opens new ways for exploiting the unique features of this material.

\acknowledgements

We acknowledge financial support from the European Research Council, the Royal Society, the Spanish Ministry of Economy (MINECO) through Grant no. FIS2011-23713, the European Research Council Advanced Grant (contract 290846) and from European Commission under the Graphene Flagship. contract CNECT-ICT-604391.

\bibliography{biblio}

\pagebreak

\appendix
 
\section{\large Appendix} 

\section{Low energy description of a stacking domain wall}
A bilayer with non-uniform stacking may be modelled, at low energies, by the Hamiltonian \cite{Neto:RMP09}
\begin{equation}
H=\left(
\begin{array}{cccc}
0 & e^{-i\phi}\Pi^+  & 0 &\gamma_\perp W_{BA}^*(\bm{r}) \\
e^{i\phi}\Pi  & 0 &\gamma_\perp W_{AB}^*(\bm{r}) & 0 \\
0 &\gamma_\perp W_{AB}(\bm{r}) & 0 & e^{-i\phi}\Pi^+  \\
\gamma_\perp W_{BA}(\bm{r}) & 0 & e^{i\phi}\Pi  & 0\end{array}
\right)
\label{H}
\end{equation}
where $\Pi=v_F(k_x+ik_y)$, and trigonal warping terms $\gamma_{3,4}$ have been neglected for simplicity. A stacking domain wall, or stacking soliton, oriented along $y$ and with its centre at $x=0$ is modelled by $W_{AB}(\bm{r})=W_{AB}(x)=W(x/L)$, $W_{BA}(\bm{r})=W_{BA}(x)=W(-x/L)$, where $L$ is the typical thickness of the soliton, and real function $W(x)$ (modelled phenomenologically, see next section) satisfies the boundary conditions $W(x\to-\infty)=1$ and $W(x\to\infty)=0$ (i.e. in the leads), thus describing a transition from pure $AB'$ stacking to $BA'$ stacking across a distance $L$. $L$ is the soliton's full-width-half-maximum (FWHM), defined by $W_{AB}(L/2)=\frac{1}{2}W_{AB}(0)$. The interlayer hopping at the centre of the soliton is $\gamma_\perp^\mathrm{SP}=\gamma_\perp W(0)\approx 88$ meV in our model, where $\gamma_\perp\approx 330 $ meV is the interlayer coupling of the AB'and BA' regions. Since it is a straight boundary, wave vector $k_y$ is conserved in transport.
Angle $\phi$ denotes the orientation of the underlying atomic lattice, with $\phi=0$ denoting intralayer A-B bonds oriented in the $y$ direction (shear soliton), and $\phi=\pi/2$ in the $x$ direction (tensile soliton).

\section{Model details \label{sec:Model}}

A microscopic derivation of the continuum model Eq. (\ref{H}) can be performed by assuming that around a given point $\mathbf{r}=(x,y)$ in the bilayer, the two layers are crystallographically aligned, and their stacking is uniform, given by an interlayer displacement $\delta$ in one of the three bond directions. One then computes the interlayer matrix $U_{ij}$, where $i=A,B$ and $j=A',B'$, as a sum of contributions from all sites in a bilayer with the displacement $\delta$\cite{Santos:PRL07, Koshino:PRB13}. Denoting by $t(\mathbf{R})$ the hopping between to carbon sites separated by a vector $\mathbf{R}=(x,y,z)$, and taking into account that around the Dirac point states exhibit a fast phase oscillation of the form $e^{i\mathbf{K}\cdot\mathbf{r}}$, we have
\begin{eqnarray}
U_{ij}&=&\sum_{m_1,m_2}t\left[\mathbf{d}+m_1\mathbf{A}_1+m_2\mathbf{A}_2+((i-j)a_{cc}+\delta)\mathbf{n}_{AB}\right]\nonumber\\
&&\times e^{i 2\pi(m_1-m_2)/3}\label{Uij}
\end{eqnarray}
where $\mathbf{d}=(0,0,d)$ is the vector separating the two layers, $a_{cc}=0.14$ nm is the carbon-carbon distance, the A-B bond direction is $\mathbf{n}_{AB}=(0,1,0)$, and the primitive vectors of the lattice are $\mathbf{A}_{1,2}=(\pm \sqrt{3}/2, 3/2,0)a_{cc}$. This gives a coupling that reaches a maximum $\gamma_\perp$ at integer $\delta/a_{cc}$. We define the dimensionless functions $W$ by factoring out this energy scale from $U$, 
\[
U_{ij}(\delta)=\gamma_\perp \left(\begin{array}{cc}
W(\delta) & W(\delta-a_{cc})\\
W(\delta+a_{cc}) & W(\delta)
\end{array}\right)
\]

The expression (\ref{Uij}) for $W$ may be expanded for $d\gg a_{cc}$, which yields a simple form for $W(\delta)$ involving only the leading harmonics \cite{Koshino:PRB13}
\[
W=\frac{1}{3}\left[1+2\cos\left(\frac{2\pi}{3}\frac{\delta}{a_{cc}}\right)\right]
\]
Note that in a bilayer $d/a_{cc}\sim 2.5$, which is not deep into the $d\gg a_{cc}$ limit. Moreover, delamination effects were observed \cite{Lin:NL13} in the regions with imperfect stacking, which weaken the interlayer coupling for $\delta\neq 0$ as compared to $\delta=0$. In our simulations we include both of these corrections using a higher order expansion in $d/a_{cc}$, which yields
\[
W\approx \frac{1}{3}\left[1+2\cos\left(\frac{2\pi}{3}\frac{\delta}{a_{cc}}\right)\right]e^{-\frac{1}{2}\sin^2(\frac{\pi}{3} \delta/a_{cc})/\sigma^2}
\]
for a given model parameter $\sigma$, which we choose as $\sigma=0.37$ to obtain the approximate experimental position of the differential conductance sidepeak for minimal shear solitons of thickness $\sim 9$ nm. A minimal soliton, AB/SP/BA, is modelled in the simplest way, with a $\delta(x)$ growing linearly with position $x$ from $\delta=a_{cc}$ (AB) to $\delta=1.5 a_{cc}$ (SP), to $\delta=2a_{cc}$ (BA). 

As is clear from Eq. (\ref{H}), we have made the approximation in our simulations that $U_{AA}=U_{BB}\approx 0$ inside the minimal AB/SP/BA soliton. This is a very good approximation, since in our model $|U_{AA/BB}/U_{AB/BA}|$, which is maximum in the center of the soliton, is very small, less than $|W(1.5 a_{cc})/W(0.5 a_{cc})|=0.03$. It is nevertheless important to assess whether this small $U_{AA/BB}$ correction affects the numerical results for the differential conductance.

\begin{figure}
   \centering
   \includegraphics[width=\columnwidth]{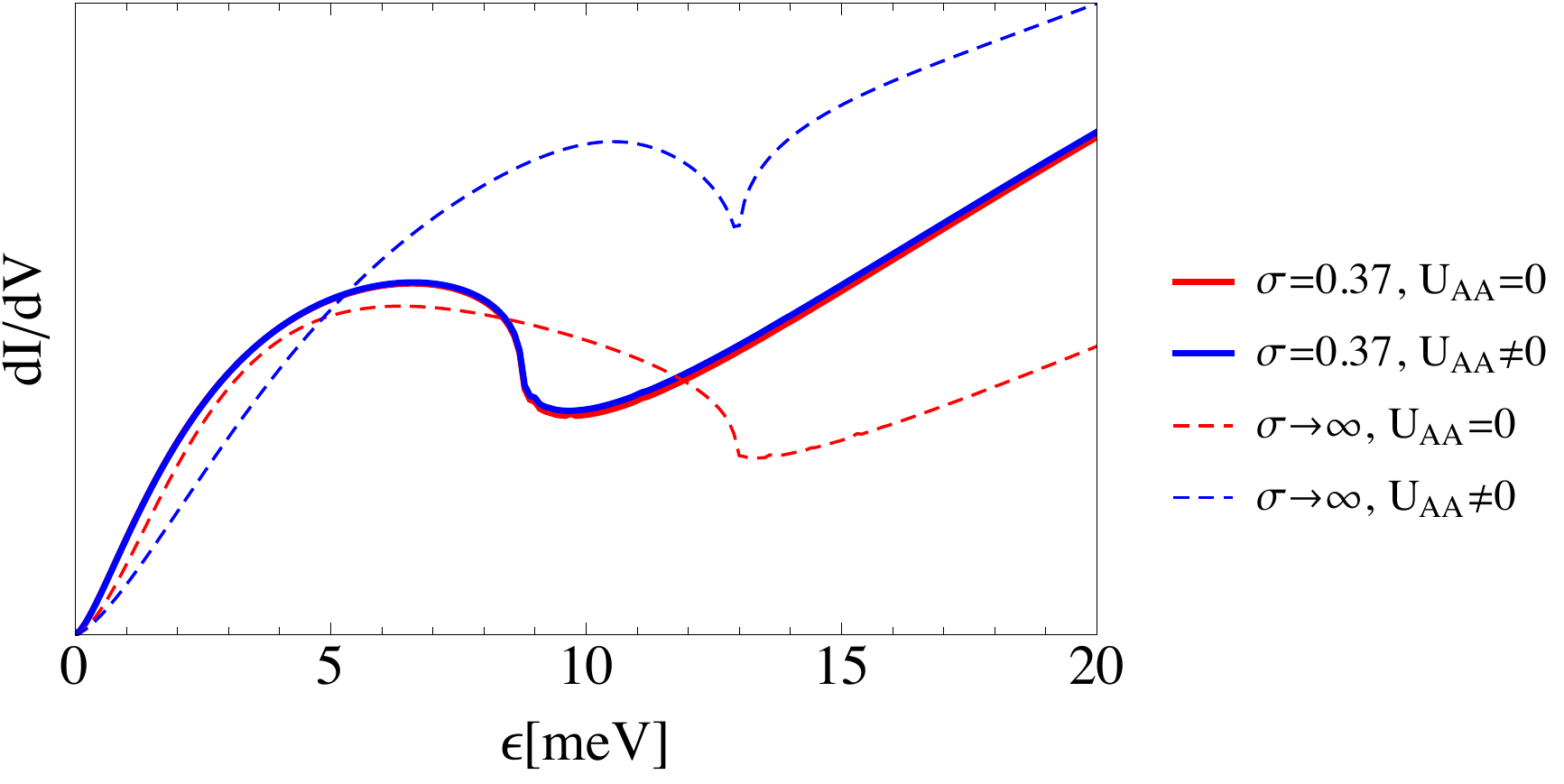} 
   \caption{Differential conductance across a 12 nm thick shear soliton, using different models. All of them exhibit similar qualitative features.}
   \label{fig:comparison}
\end{figure}

In Fig. (\ref{fig:comparison}) we show the differential conductance for a shear soliton with FWHM thickness of 12 nm, using both the $U_{AA}=0$ and $U_{AA}\neq 0$ models. We see that for $\sigma=0.37$ both curves (solid red and blue) are almost indistinguishable. We also plot  results for the leading-harmonic model ($\sigma\to\infty$, dashed lines), which also shows a (somewhat stronger) transport gap, but a different shape of the sidepeaks. In this case, the $U_{AA}=0$ (red) and $U_{AA}\neq 0$ (blue) models yield clearly different results. The qualitative structure of transport for all four models, however, is very similar.

\section{Spectrum and transport across and adiabatic stacking soliton}

\begin{figure*}
   \centering
   \includegraphics[width=\textwidth]{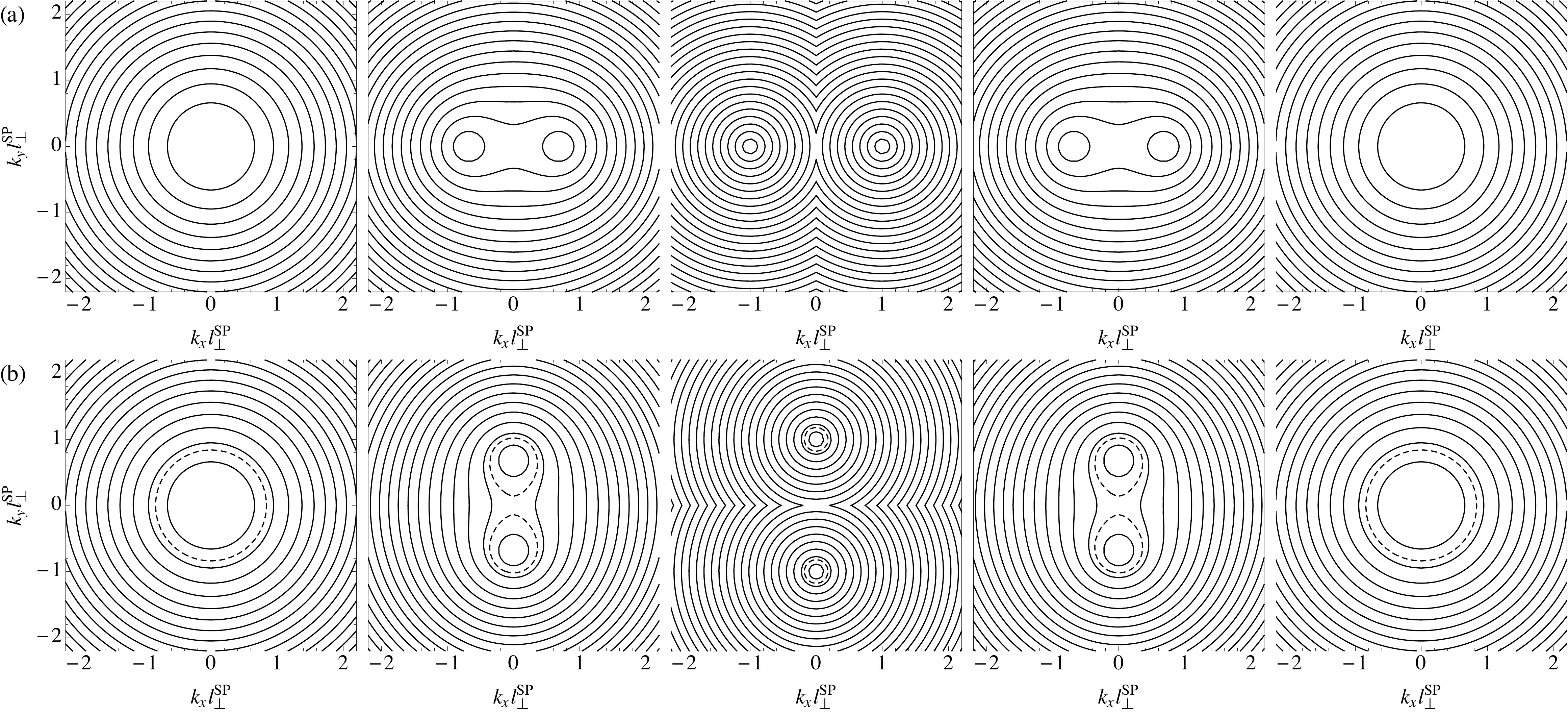} 
   \caption{Local bands at energies close to the neutrality point as one traverses a shear soliton (a) and a tensile soliton (b). The parabolic low energy bands of AB' and BA' stackings (left and right panels) evolve into a pair of Dirac cones shifted away from the K point. $l_\perp^\mathrm{SP}\approx 6.6$ nm is the interlayer coupling length. Solid contour lines correspond to multiples of 10 meV. The dashed line in (b) corresponds to energy 15.9 meV, above which adiabatic transport becomes possible.}
   \label{fig:bands}
\end{figure*}

In order to understand transport through a stacking soliton, it is important to grasp the evolution in the local bandstructure as one moves across the soliton. In the adiabatic limit $L\to \infty$, carriers of a given energy $\epsilon$ and wavevector $k_y$ will be (perfectly) transmitted if and only if there are available states for said $\epsilon,k_y$ throughout the traversal, so the band structure gives a precise picture of transport in this limit.

To compute the low energy band structure at different points across the soliton, we fix $x=x_0$ in Eq. (\ref{H}) for various positions $x_0$, and obtain the eigenvalues closest to zero around $(k_x,k_y)=0$ (the K or K' point of the bilayer). Fig. \ref{fig:bands} shows the isoenergetics of these bands across a shear soliton (a) and a tensile soliton (b). The parabolic AB' dispersion $\epsilon_\mathrm{AB'}\approx (\hbar v_F|\bm{k}|)^2/\gamma_\perp$ (left panels) evolves into two Dirac cones $\epsilon_\mathrm{SP}\approx \hbar v_F|\bm{k}\mp\bm{k}_\mathrm{SP}|$, shifted by a momentum $\bm{k}_\mathrm{SP}=\pm[\cos(\phi),\sin(\phi)]/l_\perp^\mathrm{SP}$, which is oriented in the $x$ direction for a shear soliton, or the $y$ direction for a tensile soliton. The interlayer coupling length is defined as $l_\perp^\mathrm{SP}=\hbar v_F/\gamma_\perp^\mathrm{SP}\approx 6.6$ nm. 

\begin{figure}
   \centering
   \includegraphics[width=\columnwidth]{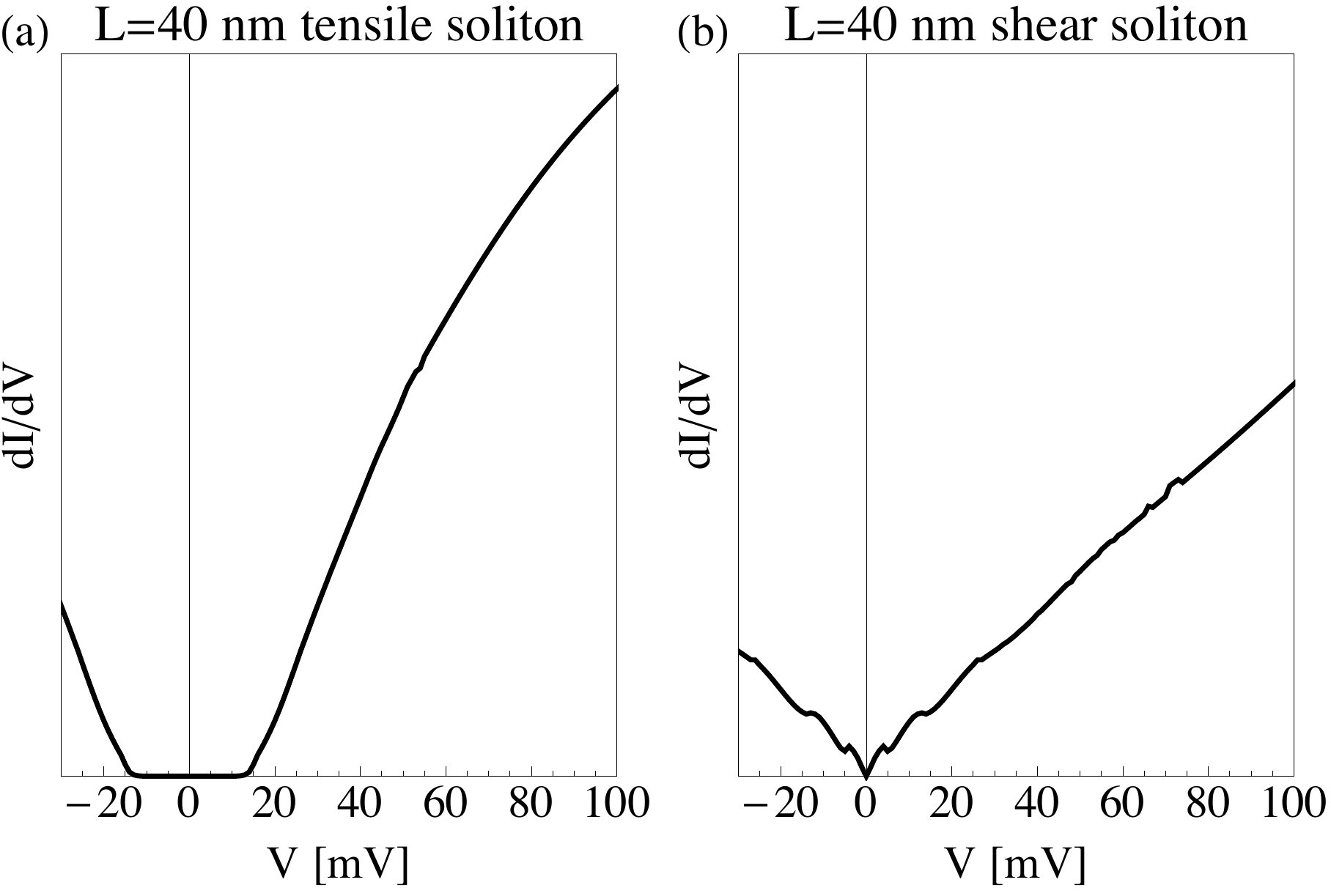} 
   \caption{Differential conductance across a very long soliton $L=40$ nm. (a) corresponds to the tensile configuration, and displays a sharp transport gap, as expected from adiabatic arguments. (b) corresponds to the shear soliton, and is gapless. The weak structure in the latter correspond to chirality-related non-adiabatic corrections.}
   \label{fig:adiabatic}
\end{figure}

It is clear from this picture that ballistic transmission through a tensile soliton will be completely suppressed in the adiabatic limit up to a finite energy, beyond which the incoming isoenergetic line overlaps with that of the shifted Dirac cones at the centre of the soliton. Equating $\epsilon_\mathrm{AB'}=\epsilon_\mathrm{SP}$ for $k_x=0$, we find that transport becomes possible around $k_yl_\perp^\mathrm{SP}\approx 0.82$, for energies above $\epsilon\approx 15.9$ meV [dashed line in Fig. \ref{fig:bands}(b)]. Above this energy, a very long tensile soliton exhibits a transmission that rises monotonously from zero, see Fig. \ref{fig:adiabatic}(a). This threshold becomes smooth at finite soliton thickness, but suppression below this energy remains strong for realistic thickness of $L=9$ nm, see main text.
Low energy transmission through an adiabatic shear soliton is starkly different, since there are always available states throughout the traversal around $\epsilon=0$ and $k_y=0$, see Fig. \ref{fig:bands}. Hence, no insulating adiabatic shear soliton should be expected, see Fig. \ref{fig:adiabatic}(b). The tensile transport gap closes smoothly as the soliton is rotated from tensile to soliton, see main text. 

The condition for the adiabatic regime is that the incoming wavelength $\lambda$ greatly exceed the soliton thickness $L$. This yields $L>200~\mathrm{nm}/\sqrt{\epsilon [\mathrm{meV}]}$, or $L>37$ nm for $\epsilon=30$ meV. Corrections beyond the adiabatic limit introduce complexity to the transport curves that cannot be accounted for simply by analysing the local band structure. The structure of the wavefunctions becomes important, and signatures of the chirality of carriers appear. The non-monotonous features of the shear case, already visible in the $L=40$ nm results of Fig. \ref{fig:adiabatic}(b) become more prominent, and develop, as $L$ is reduced, into the differential conductivity profiles discussed in the main text.

\section{Non-adiabatic transport across a shear soliton}

\begin{figure}
   \centering
   \includegraphics[width=\columnwidth]{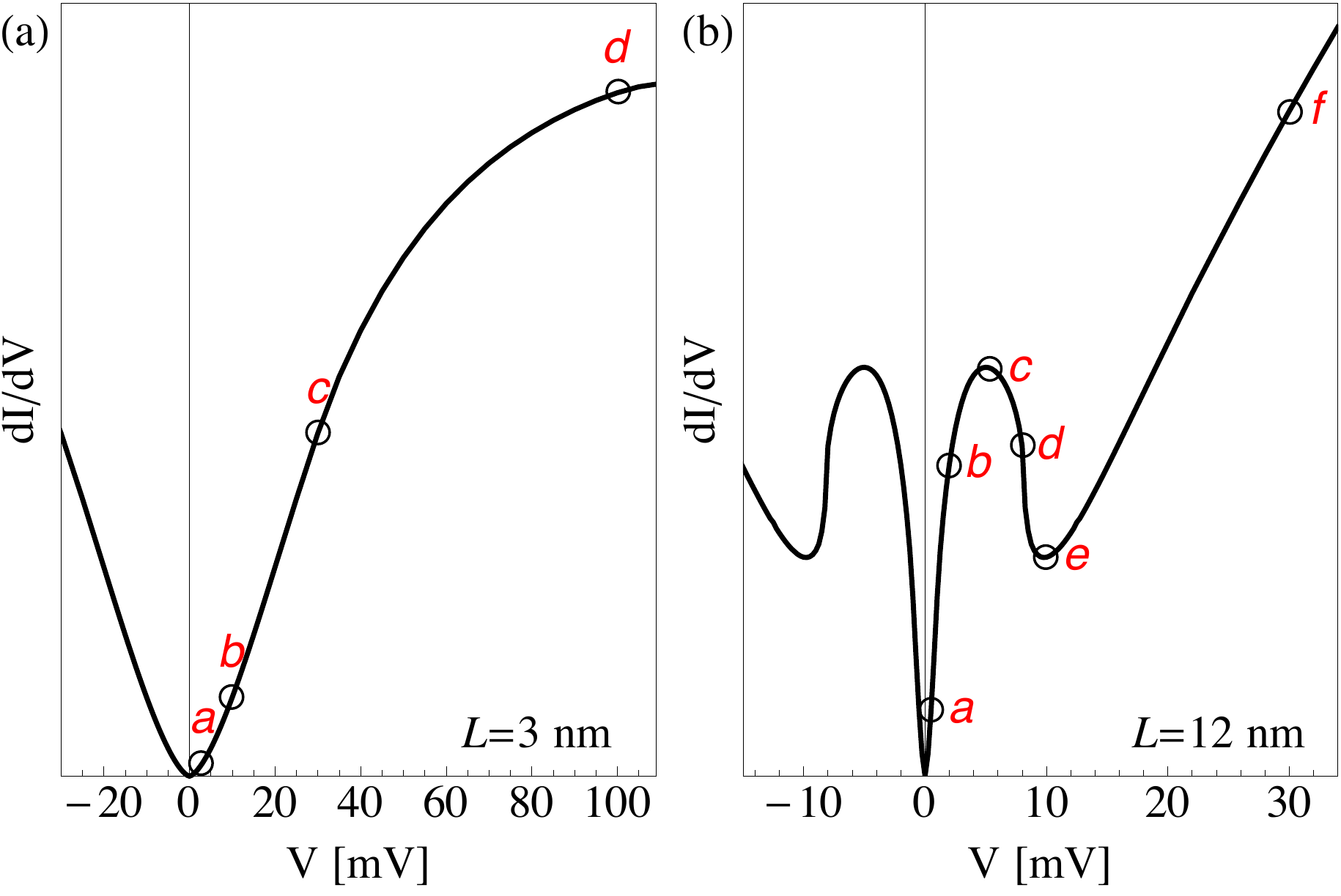} 
   \caption{Integrated transmission across a short shear soliton, $L=3~\mathrm{nm}<l_\perp^\mathrm{SP}$ (a) and a long shear soliton, $L=12~\mathrm{nm}>l_\perp^\mathrm{SP}$ (b).}
   \label{fig:TvsV}
\end{figure}

\begin{figure*}
   \centering
   \includegraphics[width=\textwidth]{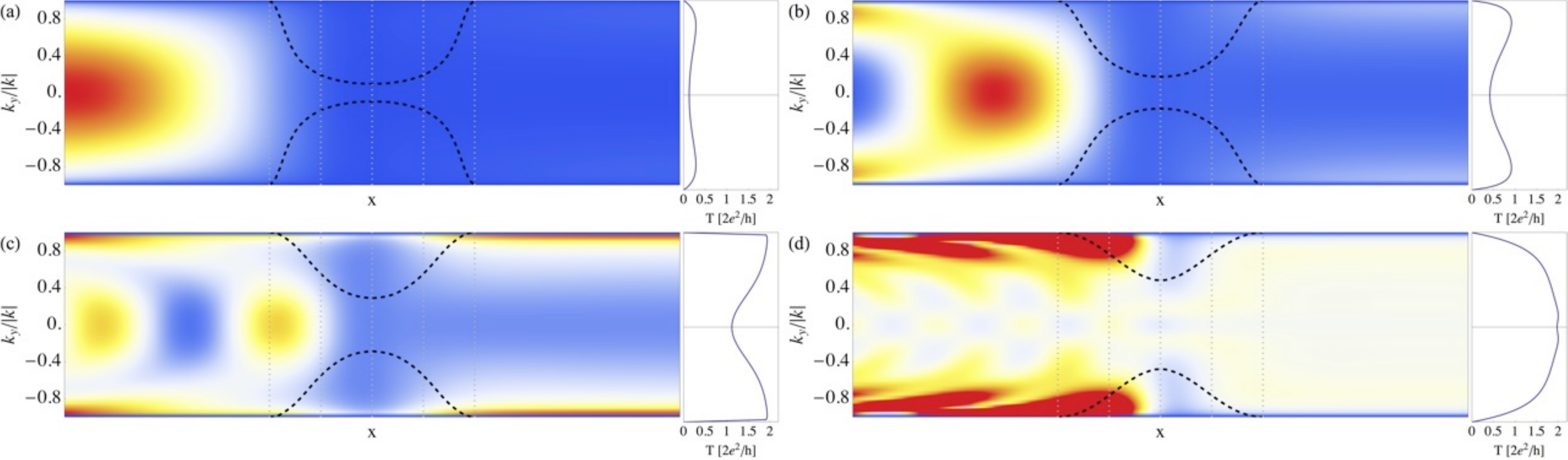} 
   \caption{Density of scattering states incident from the left as a function of  wavevector $k_y$ and position $x$ across a strongly non-adiabatic $L=3$ nm shear soliton. Red is maximum, blue is zero. Different panels correspond to different energies of the incoming state, shown as circles in Fig. \ref{fig:TvsV}(a). Vertical dotted lines are the positions with local bandtructures shown in Fig. \ref{fig:bands}(b). Black dashed lines show the window of wave vectors $|k_y|<|k_y^\mathrm{max}(x)|$ for which adiabatic transmission is allowed. The $k_y$ resolved transmission is shown by the curve to the right of each panel.}
   \label{fig:SNA}
\end{figure*}

\begin{figure*}
   \centering
   \includegraphics[width=\textwidth]{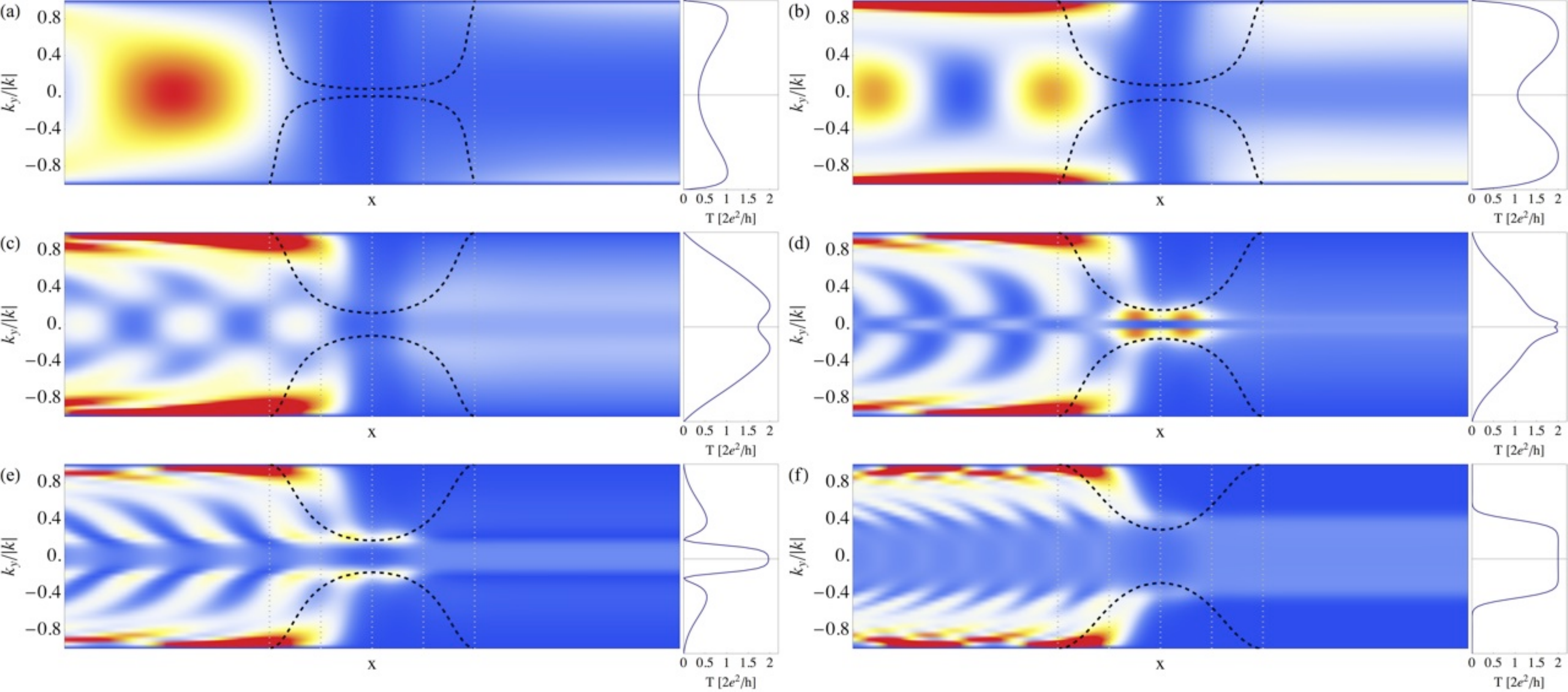} 
   \caption{The same as Fig. \ref{fig:SNA} but for a weakly non-adiabatic soliton, $L=12$ nm.}
   \label{fig:WNA}
\end{figure*}

Using the recursive Green's function algorithm described in Sec. \ref{sec:RGF} we now present and analyse in detail the transport properties across a shear soliton beyond the adiabatic approximation.  Due to the presence of the length scale $l_\perp^\mathrm{SP}\approx 6.6$ nm, in addition to the incoming wavelength $\lambda$, the non-adiabatic soliton exhibits two distinct transport regimes, termed here strongly and weakly non-adiabatic. They are readily apparent in the structure of the differential conductance (main text and Fig. 
\ref{fig:TvsV}). The differential conductance $dI/dV$ for a bias $V$ is simply proportional to the transmission $T(\epsilon=eV,k_y)$ integrated over $k_y$, $dI/dV=4 \frac{e^2}{h}W\int \frac{dk_y}{2\pi} T(eV, k_y)$. In Fig. \ref{fig:TvsV} we reproduce the essential features of the two non-adiabatic transport regimes. The first [Fig. \ref{fig:TvsV}(a)] corresponds to $L<l_\perp^\mathrm{SP}<\lambda$, and exhibit a simple monotonous $dI/dV$. This is the strongly non-adiabatic regime, that is fully independent of the electronic states inside the soliton. A perfectly abrupt (valley-conserving) interface between an AB' and a BA'-stacked bilayer, as well as a strongly-non-adiabatic tensile soliton would display the same transport properties. Due to the opposite chirality of the AB' and BA' stackings, normal incidence is suppressed. For thicker solitons, such that $l_\perp^\mathrm{SP}<L<\lambda$, a non-trivial dependence of the transmission with $k_y$ and $\epsilon$ develops, the differential conductance becomes non-monotonous [Fig. \ref{fig:TvsV}(b)] and the chiral structure of the states inside the soliton become relevant. This is the weakly non-adiabatic regime. Interestingly, it is the most relevant experimentally, since reported values for $L$ lie in the $\sim 10$ nm region.

The spatial dependence of the scattering states as a function of $k_y$ for various energy is presented, both for the strongly (Fig. \ref{fig:SNA}, $L=3$ nm) and weakly (Fig. \ref{fig:WNA}, $L=12$ nm) non-adiabatic regimes. They include also the case of high energies in which transport approaches the adiabatic limit. The different band structures of Fig. \ref{fig:bands}(a) correspond to positions marked by dotted vertical lines. The black dashed line corresponds to the maximum $k_y^\mathrm{max}(x)$ available for each energy as the soliton is traversed. The minimum of this window corresponds to the centre of the soliton, and we denote it by $k_y^\mathrm{max}(0)$.  The curve on the right of each panel is the $k_y$ resolved transmission. Each panel correspond to different energies of the incoming carrier, marked by the circles of Fig. \ref{fig:TvsV}.

We see that the strongly non-adiabatic  case, Fig.  
\ref{fig:SNA}, evolves quite simply, from a transmission dominated by grazing angles at low energies (due to the opposite chirality of the AB' and BA' regions), into a transmission that is large for all momenta $k_y$ at large energies (adiabatic limit). In contrast, in the weakly non-adiabatic limit, Fig. \ref{fig:WNA}, things are more complicated. We see that the maximum in the lowest energy transmission at grazing angles becomes shifted towards normal incidence as the energy is increased. At some point [panel (d)], the maximum lies at normal incidence, and remains there as energy is increased further. The integrated transmission however exhibits a minimum at (e), due to the appearance of a zero for a $k_y\approx k_y^\mathrm{max}(0)$. This zero moves to higher wave vectors as energy crosses this intermediate regime, characterised by $\lambda\sim l_\perp^\mathrm{SP}$. At high enough energy, however, the transmission above said zero becomes negligible, and only momenta $k_y<k_y^\mathrm{max}(0)$ are transmitted [panel (f)]. This marks the onset of the adiabatic regime $\lambda<l_\perp^\mathrm{SP}<L$, which displays an increase of the differential conductance with bias.

\section{Transport through a sequence of solitons}

\begin{figure}[htbp]
   \centering
   \includegraphics[width=\columnwidth]{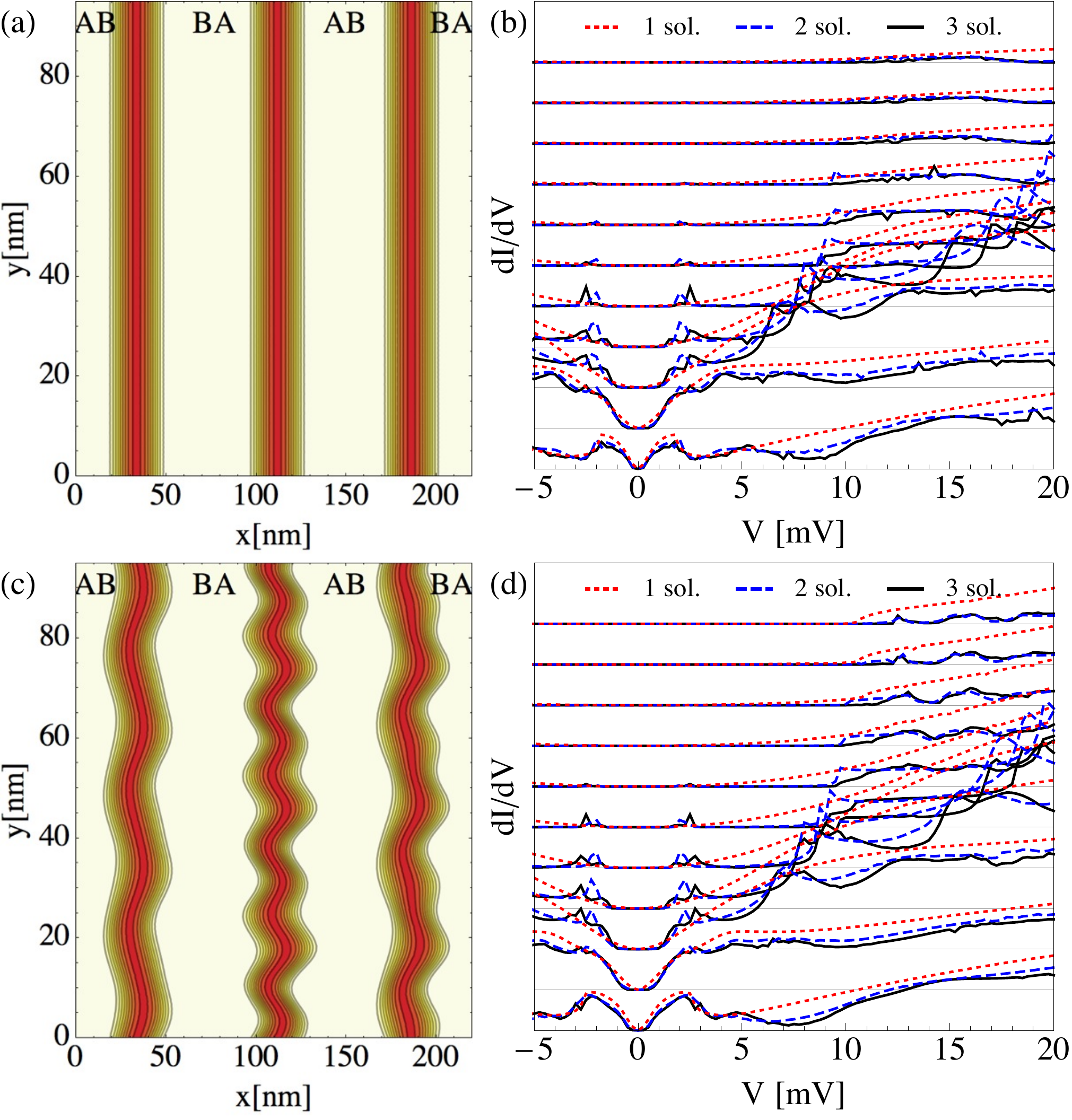}
   \caption{Differential conductance across a sequence of three solitons of thickness $\sim 9$ nm. Straight solitons (a) result in a differential conductance (b) similar to that of distorted solitons (c), shown in (d). Dotted (red), dashed (blue) and solid (black) lines correspond to transport through one, two and three solitons. Curves are offset and evolve from bottom to top by changing the orientation of the underlying bilayer lattice, i.e. the average character of the solitons, from shear (bottom) to tensile (top). This is a generalisation of the results in Fig. 2b in the main text.}
   \label{fig:threesolitons}
\end{figure}

Real samples may exhibit a number of stacking solitons. Their global configuration depends strongly on the boundary conditions of the bilayer. For twisted bilayers, solitons will form a triangular network, with nodes of $AA$ stacking \cite{Alden:PNAS13}. In the case of negligible relative rotation of the two layers in the boundary, open solitons will not cross. They will tend to be parallel at long distances, but may curve due to local distortions. 

A relevant question for the present work is whether the transport features predicted with a single straight soliton across the sample hold in the presence of a number of possibly curved solitons. In this section we present a simulation that shows this is indeed the case. We build a configuration of one, two and three parallel solitons of a thickness around 9nm, placed around 80 nm apart. We consider both straight solitons and a random distortion to the profile and thickness of each soliton, which locally changes their character between shear and tensile. The distortion is extended periodically, with a period of $W=200$ nm, so that once more we may compute the transmission $T(\epsilon,k_y)$, where $k_y$ is now in the Brillouin zone of the distortion. This allows for an analogous, though more costly, computation of the integrated differential conductance to that of the main text.

The results are presented in Fig. \ref{fig:threesolitons}. In panels (a,b) we show the spatial configuration of the three straight solitons, and the corresponding differential conductance, respectively. Different curves in (b) correspond to configurations with only the leftmost soliton (dotted red), the two leftmost (dashed blue), and the three solitons (solid black) in (a). Curves evolve from bottom to top by changing the character of the solitons, from pure shear to pure tensile. We see that the addition of more soliton barriers reinforces the transport gap, and moreover, does not wash out the side peak feature of the weakly non-adiabatic shear solitons. In fact, this feature becomes more marked with the addition of a sequence of solitons, and emerges also for a sequence of solitons that are not of the purely shear type. Panel (d) shows results analogous to (b), but with a random distortion performed on the three solitons, as shown in (c). These distortions locally perturb the thickness and the orientation of each soliton, making them non-parallel and non-uniform at scales of around 30 nm. We see that at low energies this has a negligible effect on transport. The reason is that transmission only depends strongly on transverse momentum $k_y$ at energies above some meV, so that once it is integrated over all momental, differential conductance is not strongly affected by soliton misalignment, which is roughly equivalent to a shift in transverse momentum at each soliton.

We therefore see that transport is not modified qualitatively by sequential scattering on distorted solitons. Indeed, the transport gap plus side-peak features resemble the experimental observations even more closely in this case. The transport gap acquires a smoother U-shaped profile. This suggests that the conclusions drawn in the main text regarding transport through a single straight soliton are applicable to more complex soliton configurations expected in realistic bilayers. This should include also non-parallel soliton networks, as those arising from a slight interlayer rotation, at least at low energies. The argument is that a soliton misalignment, as found in soliton networks, is locally similar to the soliton distortions depicted in Fig. \ref{fig:threesolitons}(c), and is expected to exhibit similar conductance.

\section{Magnetotransport across a soliton}

\begin{figure}
   \centering
   \includegraphics[width=\columnwidth]{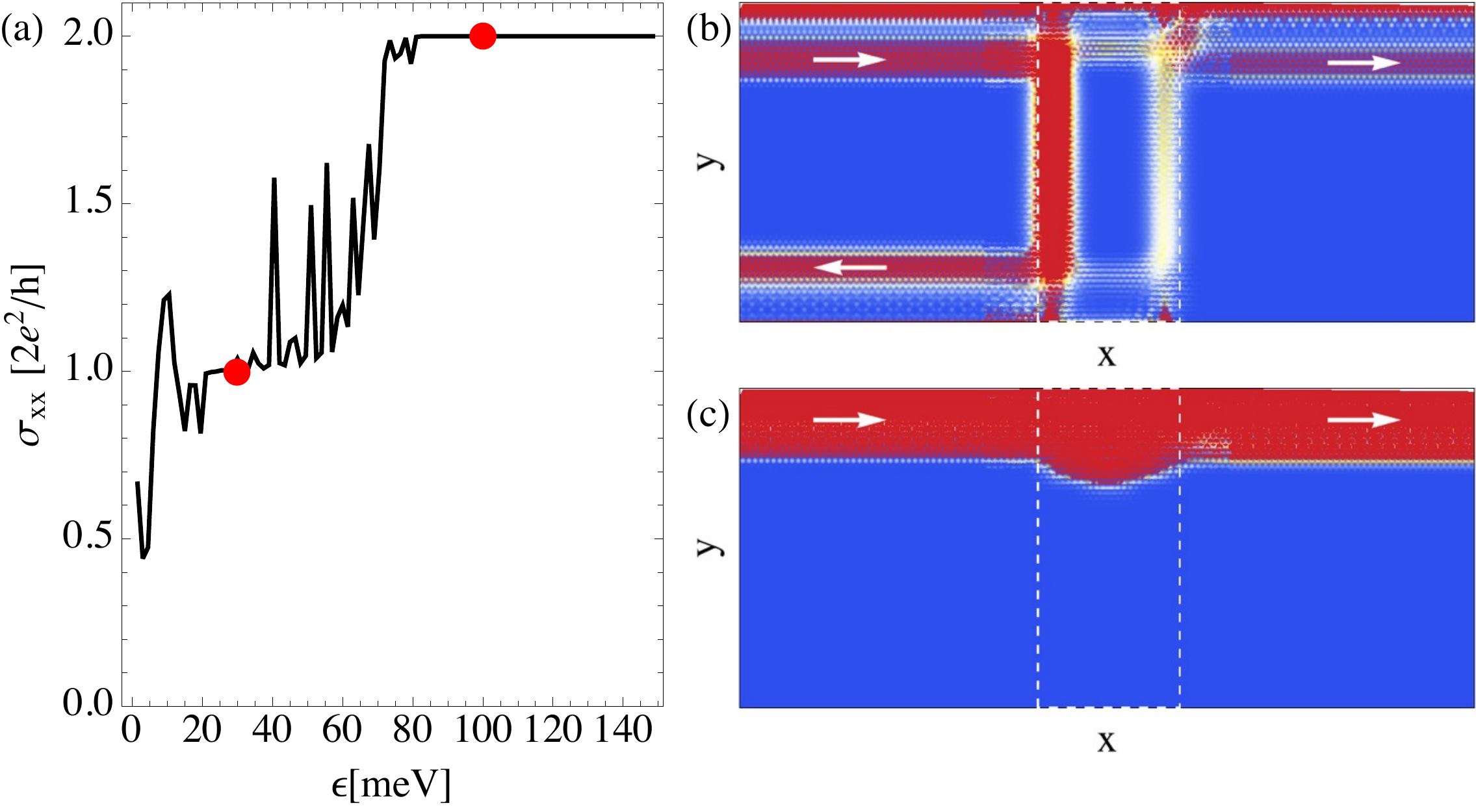} 
   \caption{(a) Conductance of a two terminal bilayer graphene Hall bar as a function of the energy of carriers in the Hall regime. A shear soliton lies across the Hall bar, marked by dashed lines in (b,c). Conductance quantisation is destroyed below a threshold energy due to inter-edge scattering along the soliton. Spatial density of scattering states at the energies 30 meV and 100 meV are shown in (a) and (b) respectively (blue is zero density, red is maximum). These correspond to the non-adiabatic and adiabatic magneto transport regimes [red dots in (a)]. Current flow is indicated by arrows.}
   \label{fig:MS}
\end{figure}

\begin{figure}
   \centering
   \includegraphics[width=\columnwidth]{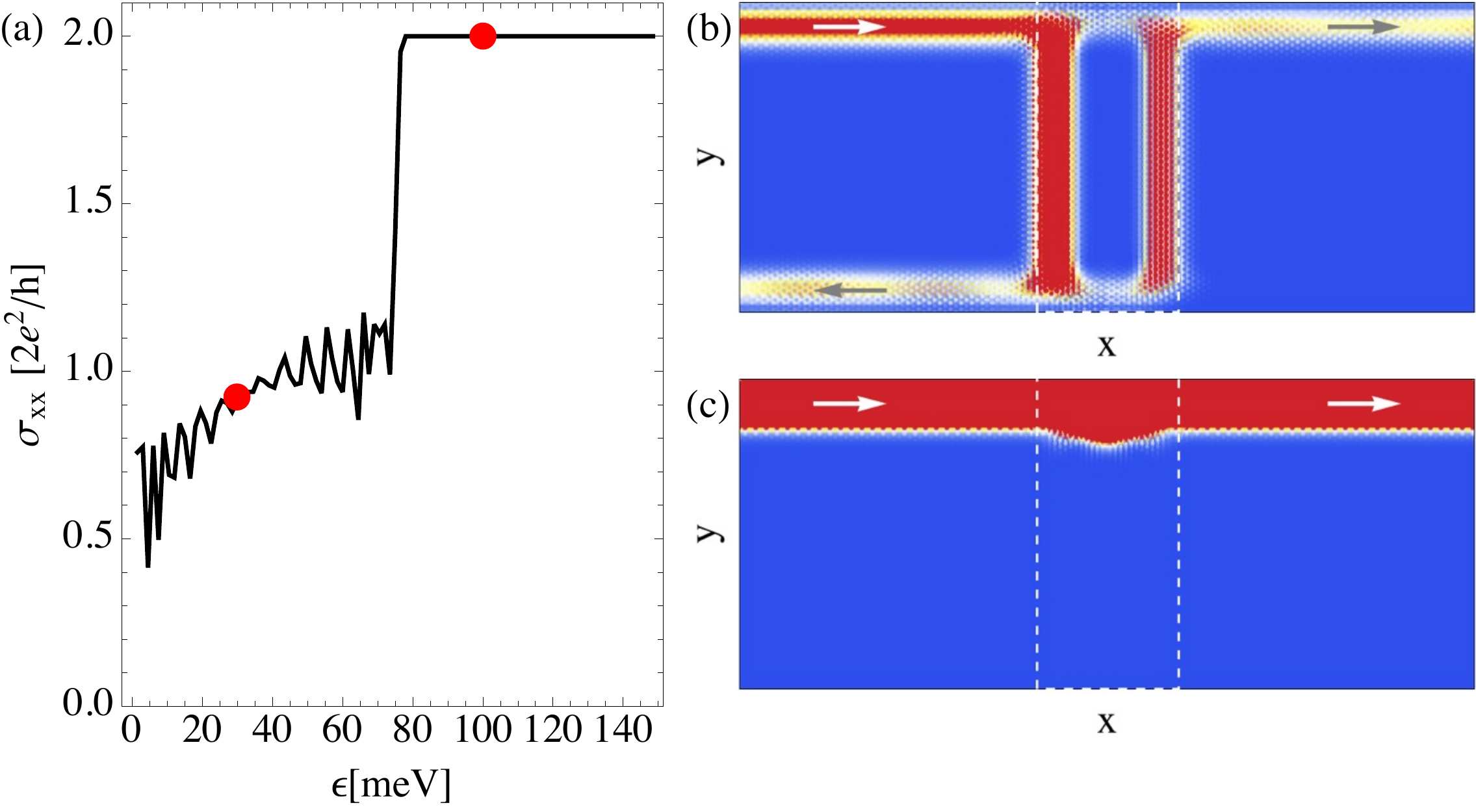} 
   \caption{Same as Fig. \ref{fig:MS} for a tensile soliton.}
   \label{fig:MT}
\end{figure}

The possibility of a stacking soliton to effectively connect at low energies the opposite edges of a bilayer graphene Hall bar was presented in the main text. Here we present the curves of the magneto conductance $\sigma_{xx}$ in a two terminal Hall bar as a function of energy. The soliton across the Hall bar is assumed thinner than $l_\perp^\mathrm{SP}$ for simplicity. As the energy of the incoming edge mode is increased, its wavelength crosses the value of the soliton thickness. A dramatic change occurs at such energy. Below it, conductance is non-quantised, and fluctuates around approximately $2e^2/h$, due to effective inter edge backscattering along the soliton. Above it, the soliton becomes adiabatic to the incoming electrons, and inter-edge scattering becomes very strongly suppressed. In this $L<l_\perp^\mathrm{SP}$ regime, a shear and a tensile soliton exhibit the same behaviour. Fig. \ref{fig:MS} shows the results for a shear soliton and Fig. \ref{fig:MT} for a tensile soliton.

\section{Recursive Green's functions and transport across a stacking soliton}
\label{sec:RGF}

In this section we discuss a flexible method to compute the transport properties of a bilayer described by Eq. (\ref{H}). $k_y$ is a good quantum. If the sample width is effectively infinite, we can ignore boundaries in the $y$ direction. For each $k_y$ we have a 1D problem, with total transmission given by Caroli's formula \cite{Caroli:JPCSSP71,Jauho:96} in terms of the Green function between $x=0$ and $x=L$
\[
T=g_s\mathrm{Tr}\left[\Gamma_L G^r_{L0}\Gamma_0^\dagger G^a_{0L}\right]
\]
where $\Gamma_{x}=\frac{i}{2}(\Sigma_{xx}-\Sigma_{xx}^\dagger)$ are inverse lifetimes, and $G^r_{x'x}=\left(G^a_{xx'}\right)^\dagger$ is the retarded Green function between points $x$ and $x'$ in the 1D problem, respectively. Likewise, $\Sigma_{00}$ is the retarded self-energy due to the left lead at site $x_0=0$, and $\Sigma_{LL}$ is the retarded self energy from the right lead at site $x_N=L$. Since $H$ is a $4\times 4$ matrix, so is $G$, $\Sigma$ and $\Gamma$. 

Note that the $g_s=2$ above accounts for the spin degree of freedom, assumed degenerate here. The valley degree of freedom is not explicitly included, since the discretisation procedure, to be discussed below, includes a "spurious" doubling of the propagating states in the leads, as a result of the no-go theorem. Thus, an additional valley degeneracy factor $g_v=2$ is effectively hidden inside the trace.

To compute the $G$ and $\Gamma$ matrices, we proceed by discretising the 1D problem into sites at position $x_n=n a$, where $a=0.14$ nm is the carbon-carbon distance in graphene. At each site, we consider four degrees of freedom (orbitals), representing the $A, B, A', B'$ p-orbitals of the bilayer's unit cell. Thus, we can turn the continuum model back into a discrete model. The result, however, is more economical than the original atomic tight binding, since each site has always 4 orbitals for any value of $\phi$. In the atomic lattice, a given (conmensurate) $\phi$ would correspond to a wide unit cell in the 1D problem, as in nanotubes of arbitrary chirality.

In the discrete lattice, the momentum operator takes the form of a finite difference $k_x a\approx \frac{i}{2}|x_{n+1}\rangle\langle x_{n}|$+h.c. Thus, the Hamiltonian $H$ can be written in the form of a tight binding chain, with an onsite $\mathcal{H}_0$ and nearest neighbour hopping $\mathcal{V}$,
\[
H=\sum_n \mathcal{H}_0|x_n\rangle\langle x_n|+ \mathcal{V}|x_{n+1}\rangle\langle x_n|+\mathcal{V}^\dagger|x_{n}\rangle\langle x_{n+1}|
\]
where
\begin{eqnarray*}
\mathcal{H}_0&=&\left(
\begin{array}{cccc}
0 & -ie^{-i\phi}v_Fk_y  & 0 &\gamma_\perp W_{BA}(x) \\
ie^{i\phi}v_Fk_y  & 0 &\gamma_\perp W_{AB}(x) & 0 \\
0 &\gamma_\perp W_{AB}(x) & 0 & -ie^{-i\phi}v_Fk_y  \\
\gamma_\perp W_{BA}(x) & 0 & ie^{i\phi}v_Fk_y  & 0\end{array}
\right)\\
\mathcal{V}&=&\left(
\begin{array}{cccc}
0 & i\frac{3}{4}\gamma_0e^{-i\phi}  & 0 &0 \\
i\frac{3}{4}\gamma_0e^{i\phi}  & 0 &0 & 0 \\
0 &0 & 0 & i\frac{3}{4}\gamma_0e^{-i\phi}  \\
0 & 0 & i\frac{3}{4}\gamma_0e^{i\phi}  & 0\end{array}
\right)
\end{eqnarray*}
where we have used $v_F=\frac{3}{2}\gamma_0 a$, and $\hbar=1$. Note that matrix $\mathcal{V}$ is not hermitian.

We next compute the self energy from the right and left lead, which is given by $\Sigma_{00}=\mathcal{V} g_L \mathcal{V}^\dagger$ and $\Sigma_{LL}=\mathcal{V}^\dagger g_R \mathcal{V}$, where $g_{L,R}$ are the retarded surface Green's functions of the left and right lead respectively, when decoupled from the central region $0<x<L$ containing the soliton. They satisfy the consistent Dyson equation
\begin{eqnarray*}
g_L(\omega)&=&(\omega-\mathcal{H}_0)^{-1}-\mathcal{V} g_L(\omega) \mathcal{V}^\dagger\\
g_R(\omega)&=&(\omega-\mathcal{H}_0)^{-1}-\mathcal{V}^\dagger g_R(\omega) \mathcal{V}\\
\end{eqnarray*}
These equations can be solved in a number of ways, the simplest (though not the most efficient) is iteration, using a small positive imaginary part in $\omega$ to ensure convergence towards the retarded solution.

Finally, we compute the retarded propagator between left and right leads, $G^r_{L0}$, and from it, the advanced $G^{a}_{0L}=(G^r_{L0})^\dagger$, using the recursive Green's function method. It consists in the iterative application of the Dyson equation to obtain by recursion the propagator $G^{(m)}_{m0}$ between the endpoints $x_0$ and $x_m$ of a portion of the system lattice, consisting of sites $n=0\dots m$. The $G^{(m)}_{m0}$ are computed in the presence of the leads, that enter as self-energies $\Sigma_{00}$ and $\Sigma_{LL}$ into the first and last sites, respectively. The last iteration, therefore, yields the desired $G^r_{L0}=G^{(N)}_{N0}$. The recursive relations read
\begin{eqnarray*}
G^{(m+1)}_{m+1,0}&=&G^{(m+1)}_{m+1,m+1}\mathcal{V}G^{(m)}_{m,0}\\
G^{(m+1)}_{m+1,m+1}&=&(\omega-\mathcal{H}_0-\mathcal{V}G^{(m)}_{m,m}\mathcal{V}^\dagger)^{-1}
\end{eqnarray*}
As seed, we need to set $G^{(0)}_{00}=(\omega-\mathcal{H}_0-\Sigma_{00})^{-1}$, and we must add the self energy from the right lead $\Sigma_{LL}$ to $\mathcal{H}_0$, upon addition of the last site $m=N$ at position $x_N=L$.

\end{document}